%% file: main.tex
\documentclass[fleqn,10pt]{wlscirep}
\usepackage[utf8]{inputenc}
\usepackage[T1]{fontenc}
\usepackage{bm}
\usepackage{tabularx}
\usepackage{booktabs}
\title{High-temperature operation of III-nitride high-electron-mobility transistors}

\author[1]{Yi-Chen Liu}
\author[1]{Jacklyn Zhu}
\author[2]{John Niroula}
\author[2]{Hridibrata Pal}
\author[2]{Tomás Palacios}
\author[1,*]{Savannah R. Eisner}
\affil[1]{Department of Electrical Engineering, Columbia University, New York, NY 10027 USA}
\affil[2]{Microsystems Technology Laboratories, Massachusetts Institute of Technology, Cambridge, MA 02139 USA}

\affil[*]{e-mail: savannah.eisner@columbia.edu}

\begin{abstract}
High-electron-mobility transistors (HEMTs) made with III-nitride materials are of potential use in high-temperature electronic applications including power electronics, communications, aerospace and space exploration. However, the demands of such applications make it essential to understand the thermal limits and performance evolution of III-nitride HEMTs. Here, we analyze the high-temperature operation of III-nitride HEMTs, examining the impact on material properties, device structure, and circuit-level behavior. We explore the role of critical device layers — including barrier and channel engineering, substrate selection, and passivation strategies — in mitigating high-temperature-induced effects, and evaluate the thermal stability of III-nitride HEMTs in logic, radiofrequency, and power electronics applications. We also highlight key remaining challenges in the design and optimization of III-nitride devices for high-temperature applications.
\end{abstract}
\begin{document}

\flushbottom
\maketitle

\thispagestyle{empty}

\begin{NoHyper}
\begingroup
  \renewcommand{\cite}[1]{}
  \nocite{Applications2023, Applications2024, Roccaforte2022sic, Huang2022sic, Applications2015, Hypersonic[14], zhang2025wide, normann2006first, guardian2024geothermal, Applications2015, serp2014molten, valentian2022venus, Applications2023, Radiationcool, bbc_haber, jdpower_exhaust, Medjdoub2006, 2012(2), 2016(1), 2019(1), 2024p, 2022Y, 2024(1), 2024K, 2023(1), 2019B, 2015(4), 2016(1), Liu2021environment, 2015F, lee2015temperature, ambacher2000two, 2024Simulation, 2006, mazumdar2019nanocrack, 2006, 2009(1), Alpert2020SensitivityTemperatures, 2024ScAlN, hardy2017epitaxial, 2024ScAlN, AlN1,AlN2, 2010I,2014, 2019(1), 2019(1), 2012(1), 2010I, 2002(2),2008(1), 2002(2), 2024K, 2021(1), 2024(5), 2010(2), 2008(2), 2010, 2010(3), 2017, 2004, 2006, 2016(1), 2025J, 2024O, 2021epi, 2006(2), 2008(1), 2017inaln, 2025investigation, 2023(1), 2017inaln, 2016, 2017(4), 2021(6), 2024J, 2024m, Macron2011, 2017(4), 2024m, islam2024effect, 2024m, 2017(1), 1996, 2021(6), 2010(3), 2025investigation, 2010(3), 2004, 2000, yuan2023enhancement, 2025pGaN800, Herfurth2013, 2019(3), 2015F, 2002(2), 2023(4), 2021(8), 2019B, 2023(4), wide_bandgap, 2010, 1996bulk, 2006(2),1999, 2024K, 2016(1), 2007(1), 2023(8), 2015(1), 2023(1), 2024(1), 2007(1), 2016(1), 2022(5), 2017(4), 2016(1), 2023(8), 2023(1), 2007_for_SS, 2019(1), 2023(1), 2024(1), 2024(1), 2008(2), 2010(1), Macron2011, 2015(6), 2019(1), islam2024effect, 2022(6), 2022(6), 2007(1), 2023(4), 2024(1), 2023(1), 2019C, 2012(2), 2022(6), 2007(1), 2023(4), 2023Towards, 2022gan, 2022(6), 2007(1), 2023(4), 2022gan, 2023Towards, 2022gan, 2024pGaN, 2024HOLE2, 2020A, 2016CMOS, cuerdo2009high, ottaviani2020evaluation, akita2001high, xue2024measurement, cui2022scaling, cuerdo2009high, akita2001high, 2025bootstrap, 2019monolithic, 2016highddd, 2022optimization, newaddedIrOx, sIc2024review}
\endgroup
\end{NoHyper}

\section*{Introduction} 

Wide-bandgap semiconductors such as silicon carbide (SiC) and gallium nitride (GaN) are critical for high-temperature electronics. The intrinsic carrier concentration in semiconductors increases exponentially with temperature, meaning that narrow bandgap semiconductors like silicon (E\textsubscript{g} = 1.11 eV) typically become unsuitable above 125$^\circ$C.\cite{Applications2023, Applications2024} Dielectric isolation, such as in silicon-on-insulator technology, can mitigate leakage currents and bring the operating limit up to 300$^\circ$C. Other conventional semiconductors like germanium (E\textsubscript{g} = 0.67 eV), gallium arsenide (E\textsubscript{g} = 1.43 eV), and indium phosphide (E\textsubscript{g} = 1.35 eV) are similarly constrained. In contrast, GaN (E\textsubscript{g} = 3.4 eV) and its III-nitride (III-N) variants offer superior thermal and electronic transport properties, both at room and high temperatures. GaN is also piezoelectric, has a direct bandgap, and can be grown heteroepitaxially on cost-effective substrates.

The most widely developed GaN device architecture is the high-electron-mobility transistor (HEMT). HEMTs exploit the formation of a two-dimensional electron gas (2DEG) at a heterojunction interface to provide high-speed, high-power operation with minimal leakage even at high temperatures. Unlike doped bulk conduction in other wide-bandgap devices, such as SiC or Ga$_2$O$_3$, the polarization-based 2DEG allows GaN HEMTs to achieve high carrier density without relying on thermally activated dopants, which can suffer from compensation and require high processing temperatures.\cite{Roccaforte2022sic, Huang2022sic} GaN HEMTs were first commercialized for use in radiofrequency power amplifiers, but they now support a range of mission-critical applications where conventional silicon electronics will fail. These include high-temperature analog front-ends for radar and telemetry, logic elements for onboard control and decision-making, and integrated sensor interfaces for magnetic field detection, infrared spectroscopy, and gas monitoring in harsh environments. In power electronics, GaN HEMTs are used for efficient d.c./d.c. and d.c./a.c. conversion with low cooling overheads, enabling lighter and more compact systems in energy and transportation systems.

\begin{figure}[h]
\centering
\includegraphics[width=\linewidth]{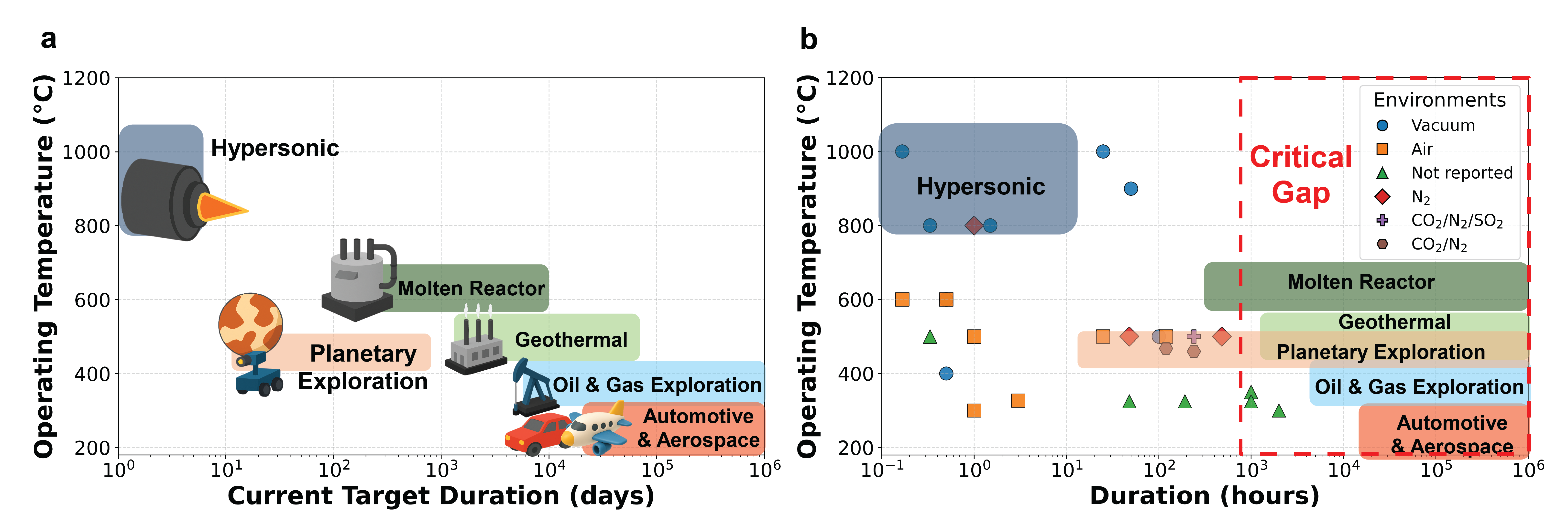}
\caption{\textbf{Temperature and duration requirements for extreme-environment applications and comparison with demonstrated III-N HEMT operating limits a) Operating temperature and target duration ranges for major application domains requiring extreme-environment electronics.} Each region represents the approximate mission envelope for a specific environment: These domains illustrate the wide span of temperature–duration requirements motivating the development of high-temperature III-nitride electronics.\cite{Hypersonic[14], guardian2024geothermal, Applications2015, serp2014molten, bbc_haber, jdpower_exhaust} \textbf{b) Reported high-temperature operating demonstrations of III-N HEMTs plotted as a function of operating temperature and sustained duration.} A shaded region labeled “Critical Gap” highlights the lack of long-duration (>10³ h) demonstrations above ~400–500 °C. Application envelopes from panel a are overlaid for comparison, showing that current device demonstrations fall short of durations required for high-temperature mission scenarios.\cite{Medjdoub2006, 2012(2), 2010(3), 1999, 2009(1), Herfurth2013, 2024(1), 2023(1), 2016(1), 2021(6), 2000, 2019(3), 2024(5), 2023(4), 2006(2), 2017(1), 2010(1), Macron2011, 2017(4), 2008(2), 2024J, Yuan2022, 2024K, yuan2023enhancement}}
\label{applications}
\end{figure}

The applications of high-temperature III-N HEMTs can be broadly categorized into two regimes: those that involve brief exposure to ultra-high temperatures, and those that involve prolonged operation at high temperatures \cite{Applications2015} (\textbf{Fig. \ref{applications}a}). The first regime is exemplified by hypersonic systems, where speeds can exceed Mach 5. In low Earth orbit, and even with thermal shielding, surface temperatures can reach 2,000$^\circ$C and the internal electronics environments over 500$^\circ$C. Therefore, hypersonic flight requires integrated sensors and high-frequency readout electronics that are capable of sustained operation above 800$^\circ$C.\cite{Hypersonic[14]} The second regime includes terrestrial and space-based technologies. In automotives, positioning control electronics located close to the engine and gearbox can improve ignition timing, fuel efficiency, and reliability, while electric and hybrid vehicles require inverters, converters, and sensors that can operate at 200$^\circ$C and last the vehicle’s lifespan. Reducing the footprint of a vehicle’s thermal management, which consumes up to one-third of their volume, also lowers carbon emissions.\cite{zhang2025wide} Aerospace systems similarly benefit from placing electronics closer to hot zones to reduce latency and mass while ensuring reliability through repeated thermal cycling.\cite{normann2006first}

In the energy sector, III-N HEMTs are essential for electronics in high temperature environments for geothermal energy extraction,\cite{guardian2024geothermal} oil drilling and logging,\cite{Applications2015} and nuclear systems such as molten salt reactors.\cite{serp2014molten} Down-hole electronics used in geo-steering and oil well logging must operate continuously for five years under temperature above 250$^\circ$C, in situations where the cost of equipment failure and production stoppages can approach millions of dollars per day. In molten salt reactors (700$^\circ$C), HEMTs support radiation-hardened sensors that monitor heat transfer loops that are critical for safety and control. HEMTs can also be used for electronics in planetary missions; exploration of Venus (460–480$^\circ$C),\cite{valentian2022venus} Mercury (430$^\circ$C), or the atmosphere of Jupiter (around 400$^\circ$C)\cite{Applications2023} require rugged sensing, acquisition, and control electronics for long-duration operations with limited cooling, high radiation exposure, and severe launch and landing shocks. As radiative dissipation scales with the fourth power of temperature,\cite{Radiationcool} electronics that can withstand higher operating temperatures permit smaller and lighter radiators for cooling systems. In industrial environments, the thermal stability of III-N HEMTs supports high-accuracy chemical sensing and control. Critical real-time feedback is needed in catalytic processes like the Haber process (450$^\circ$C),\cite{bbc_haber} automotive exhaust emissions monitoring (>530$^\circ$C),\cite{jdpower_exhaust} and sterilization validation in medical equipment.

While many III-N HEMTs have demonstrated functionality beyond 300$^\circ$C, long-term operation at high temperatures in chemically harsh or high-radiation environments remains rare. This gap is especially pronounced in applications requiring both high thermal tolerance and extended lifetimes (\textbf{Fig. \ref{applications}b}). This domain imposes stringent demands on material stability, gate control, contact reliability, and interfacial integrity. In this Review, we examine failure mechanisms in high-temperature III-N HEMTs and evaluate material, device, and circuit-level strategies to extend their operational lifetimes. We focus on GaN-based HEMTs as a model system, but many of the same principles — including polarization-driven conduction, barrier and channel engineering, passivation, and contact stability — can be broadly applied across the III-nitride family, including AlN-, InN-, and ScAlN-based devices.

\begin{figure}[h]
\centering
\includegraphics[width=1.0\linewidth]{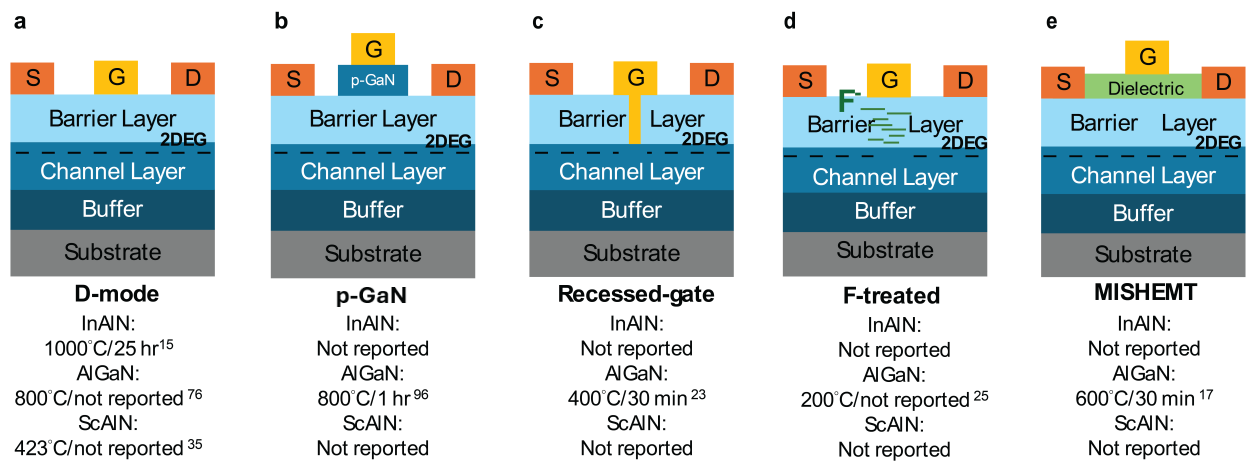}
\caption{\textbf{Comparison of III-N HEMT device structures and reported maximum operating temperatures a) D-mode HEMT.} Schematic cross-section showing a Schottky-gate depletion-mode (D-mode) HEMT, where S = source, G = gate, and D = drain. The two-dimensional electron gas (2DEG) forms at the interface between the barrier and channel layers. \textbf{b) p-GaN HEMT.} Structure incorporating a p-GaN gate layer for normally-off operation. The 2DEG resides beneath the barrier–channel interface. \textbf{c) }Recessed-gate HEMT. Gate etching reduces barrier thickness below the gate, modifying the electrostatic control of the 2DEG. 
\textbf{d) Fluorine-treated HEMT.} Negative fixed charge introduced by fluorine implantation suppresses the 2DEG under the gate, enabling normally-off operation. \textbf{e) MISHEMT.} Metal–insulator–semiconductor HEMT structure in which a dielectric layer is placed between the gate metal and barrier layer to reduce gate leakage and enhance stability.}
\label{device-types}
\end{figure}

\section*{High Temperature Operation of HEMTs}

\subsection*{Impact of Device Structure}
A key determinant of III-N HEMT thermal behavior is the engineering of the 2DEG channel and how it is modulated by gate control. This gives rise to two main operational modes: depletion-mode (D-mode, normally-on) and enhancement-mode (E-mode, normally-off) (\textbf{Fig. \ref{device-types}}). In D-mode devices, the 2DEG is naturally present at the heterointerface, allowing high current conduction without gate bias. Due to their fabrication simplicity, D-mode HEMTs are the most widely reported in high-temperature studies, up to 1000$^\circ$C \cite{Medjdoub2006, 2012(2), 2016(1), 2019(1), 2024p}. However, the need for low off-state power and fail-safe operation has accelerated interest in E-mode HEMTs, where conduction is gated and the device is naturally off at zero bias. Several strategies realizing E-mode operation include the use of a p-GaN gate epilayer to raise the conduction band and weakens quantum confinement \cite{2022Y, 2024(1), 2024K}; etched recessed-gate structures to suppress channel formation\cite{2023(1), 2019B}; and fluorine-based treatments to introduce fixed negative charge and channel pinch-off \cite{2015(4)}. Additionally, metal–insulator–semiconductor HEMTs (MISHEMTs) with gate dielectric have been explored to reduce gate leakage at temperature \cite{2016(1)}. These architectures strongly influence high-temperature reliability and must align with application-specific requirements.

\begin{figure}[ht!]
\centering
\includegraphics[width=0.65\linewidth]{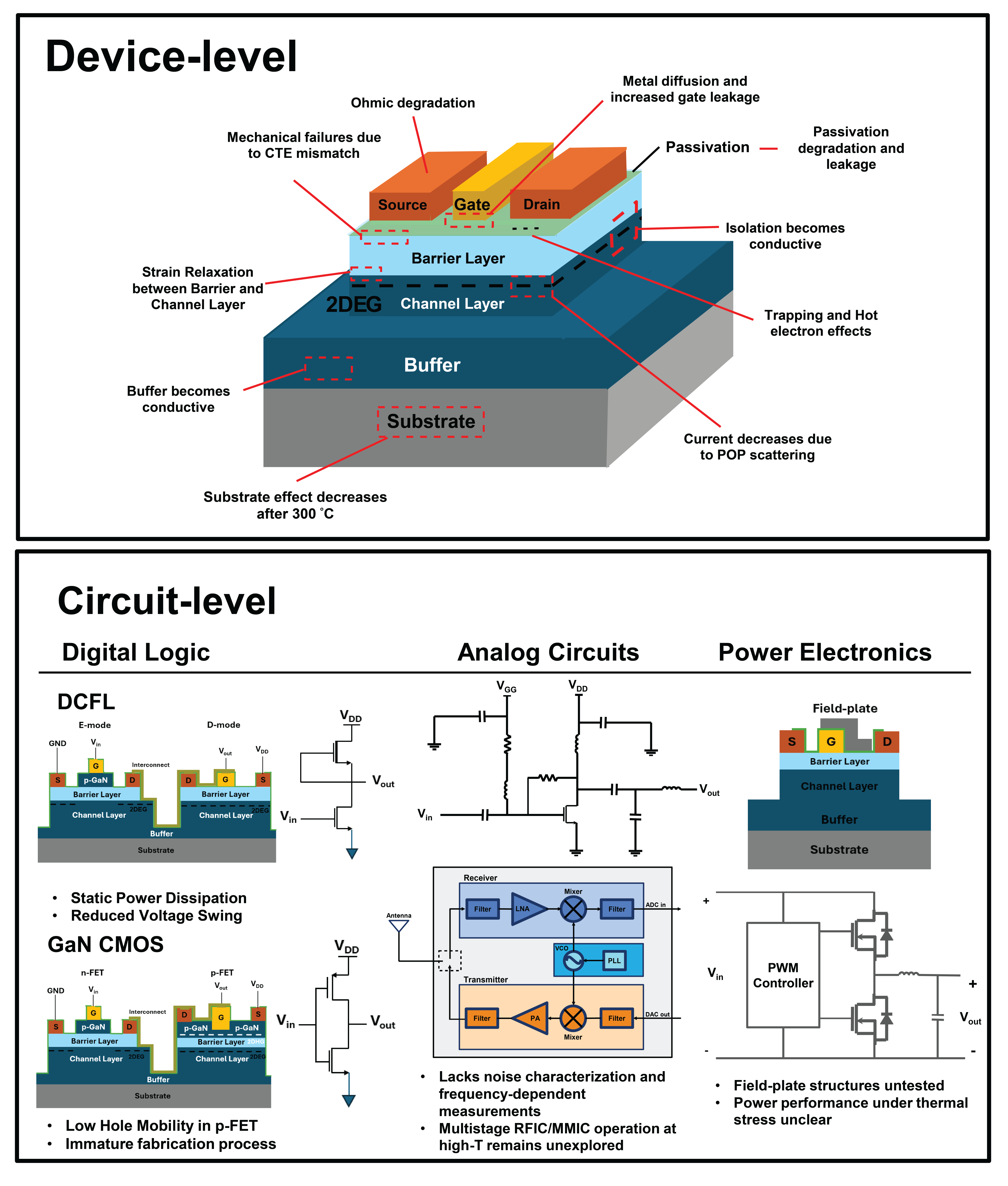}
\caption{\textbf{Device- and circuit-level failure mechanisms and performance limitations in III-N HEMT electronics a)} Device-level mechanisms. Cross-section schematic illustrating major temperature-dependent degradation pathways within the III-N HEMT structure. S, G, and D denote the source, gate, and drain contacts, respectively, and the 2DEG denotes the two-dimensional electron gas at the barrier–channel interface. \textbf{b) Circuit-level limitations.} Representative examples of circuit topologies and their high-temperature challenges. \textbf{Digital logic:} Direct-coupled FET logic (DCFL) and GaN CMOS architectures are shown; both use complementary normally-off and normally-on transistors but face constraints such as static power dissipation, reduced voltage swing, low hole mobility in p-channel devices, and immature fabrication maturity for GaN CMOS. \textbf{Analog circuits:} High-temperature analog building blocks remain limited by insufficient noise characterization, lack of frequency-dependent measurements, and the absence of validated high-temperature RF/MMIC operation. \textbf{Power electronics:} Field-plated HEMT structures (with field plate denoted F) and converter topologies illustrate outstanding questions regarding high-temperature electric-field management, device reliability under thermal stress, and unvalidated power-conversion efficiency at >300°C.}
\label{failure-circuits}
\end{figure}

The physical structure of III-N HEMTs plays a critical role in determining performance at elevated temperatures. This section examines key HEMT fabrication and design choices and their impact on device performance at temperature (\textbf{Fig. \ref{failure-circuits}}). Unless otherwise specified, all temperatures reported in this work refer to ambient temperature. Additionally, it is known that thermal stress and environmental factors (i.e. nitrogen, air, etc.) can accelerate the aging of devices.\cite{Liu2021environment} Therefore, we examine the highest reported operating temperature of different III-N HEMTs, with the understanding that a lower operating temperature or favorable environment may correspond to a longer device lifetime.

\subsubsection*{Barrier and Channel Layers}

The barrier layer, typically located above the channel in the HEMT heterostructure, induces polarization fields that control 2DEG density and mobility. Common III-nitride barrier materials include aluminum gallium nitride (Al$_x$Ga$_{1-x}$N), indium aluminum nitride (In$_x$Al$_{1-x}$N), scandium aluminum nitride (Sc$_x$Al$_{1-x}$N), and aluminum nitride (AlN). Each exhibits distinct trade-offs in lattice matching, polarization strength, ease of growth, and high-temperature reliability (Table \ref{tab:barrier_props}).

\begin{figure}[htbp]
\centering
\includegraphics[width=0.7\linewidth]{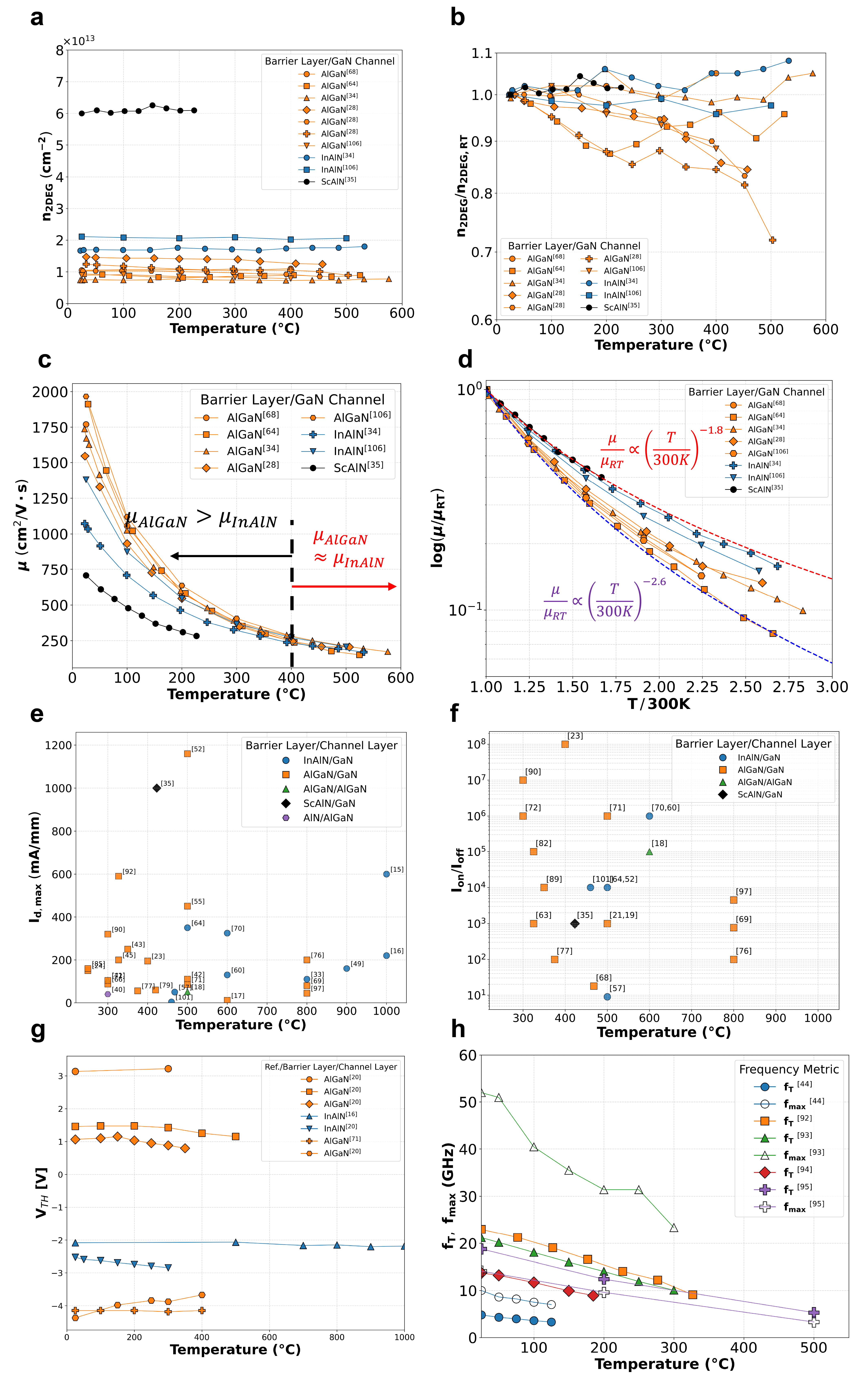}
\caption{\textbf{Temperature dependent electrical performance of III N HEMTs with different barrier and channel layers
a)} Absolute 2DEG sheet density versus temperature for AlGaN, InAlN, and ScAlN barrier devices (symbols denote reported data sets).
\textbf{b)} Sheet density normalized to room temperature (RT = 1), illustrating relative thermal stability.
c, Electron mobility $\mu$ versus temperature, showing the expected phonon-limited decrease; AlGaN-barrier devices generally retain higher $\mu$ than InAlN devices.
\textbf{d)} Mobility normalized to RT and plotted versus $T$/300 K; reference slopes ($T^{-1.8}$, $T^{-2.6}$) illustrate phonon scattering trends.
\textbf{e)} Maximum drain current $I_\text{d,max}$ versus temperature for various barrier/channel combinations.
\textbf{f)} On–off current ratio, showing increased leakage and reduced switching margin at high temperature.
\textbf{g)} Threshold voltage $V_{th}$ versus temperature, typically shifting negatively due to barrier and trap electrostatics.
\textbf{h)} High-frequency metrics (cut-off frequency $f_T$ and maximum oscillation frequency $f_\text{max}$), both decreasing with temperature due to mobility and parasitic resistance degradation.}
\label{combined}
\end{figure}

Al$_x$Ga$_{1-x}$N, the most widely used barrier, experiences thermal instability above $\sim$400$^\circ$C (\textbf{Fig. \ref{combined}c,d}), with faster degradation in on-resistance and drain current compared to InAlN and ScAlN (\textbf{Fig. \ref{combined}e}) \cite{2015F}. These effects arise from interfacial strain due to lattice mismatch with GaN, which promotes trapping, dislocation formation, and mobility degradation \cite{lee2015temperature}. Although keeping the barrier below its critical thickness (typically $\sim$30 nm for Al$_{0.25}$Ga$_{0.75}$N\cite{ambacher2000two}) preserves coherent strain during growth, high-temperature operation induces strain relaxation through thermally activated dislocation and thermal expansion mismatch \cite{2024Simulation,2006}. Relaxation via misfit dislocations is often accompanied by surface roughening and interfacial degradation \cite{mazumdar2019nanocrack}. 

In contrast, In$_x$Al$_{1-x}$N with $x \approx 0.17$ can be lattice-matched to GaN, eliminating piezoelectric polarization and strain-related degradation~\cite{2006, 2009(1)}. Its strong spontaneous polarization still enables high 2DEG densities and low sheet resistance compared to AlGaN/GaN \cite{Alpert2020SensitivityTemperatures}. However, InAlN thickness is typically limited to ensure effective gate control over the channel charge.


Sc$_x$Al$_{1-x}$N offers even stronger polarization fields and ferroelectricity \cite{2024ScAlN}, resulting in high 2DEG density (\textbf{Fig. \ref{failure-circuits}a,b}). However, it suffers from lattice mismatch with GaN at $x > 0.4$, leading to spinodal decomposition \cite{hardy2017epitaxial}. Mechanical softening of Sc-rich alloys and enhanced dislocation mobility worsen with temperature. Thus, careful control of Sc content and epitaxial conditions is essential. Sc$_{0.15}$Al$_{0.85}$N/GaN HEMTs have been demonstrated to operate up to 423$^\circ$C and exhibit high drain current density at that temperature \cite{2024ScAlN}. However, the operation duration has not been reported, and long-term reliability remains unverified.

Although less common, AlN has been explored as a barrier in ultrathin (2–3\,nm) AlN/GaN HEMTs for mm-wave applications \cite{AlN1,AlN2}. Its extreme polarization enables very high 2DEG densities, often exceeding those of ternary barriers. However, its large lattice mismatch ($\sim$2.4\%) causes strain relaxation, causing increased gate leakage and reduced electrostatic control. Moreover, the sharp interface and high fields degrade mobility due to increased scattering. While occasionally used in cap or interfacial layers to enhance carrier confinement, no high-temperature operation has been reported for these devices.

\input{table_barrier_props}

The channel layer in III-nitride HEMTs serves as the transport medium for the 2DEG. Traditionally, undoped GaN has been the standard choice due to its high electron mobility and relatively mature epitaxial quality. However, as devices are pushed to operate beyond 300$^\circ$C, GaN channels face limitations in thermally activated leakage, trap formation, and interface degradation.

Recent research has explored alternative wide-bandgap channel materials, most notably AlGaN, in structures such as AlGaN/AlGaN and AlN/AlGaN HEMTs \cite{2010I,2014}. At 300$^\circ$C, AlGaN-channel HEMTs show only half the current degradation and nearly three times less on-resistance due to alloy-scattering-limited mobility that is inherently lower and therefore less susceptible to thermal degradation \cite{2019(1)}. The wider bandgap of AlGaN also supports higher Schottky barrier heights, mitigating threshold voltage ($V_{\text{th}}$) shifts and gate leakage under thermal stress. From 350$^\circ$C to 500$^\circ$C, the dominant leakage mechanism in AlGaN channels transitions from Poole-Frenkel emission to band-to-band tunneling \cite{2019(1)}. AlGaN lacks the strong surface Fermi level pinning observed in GaN, further reducing gate leakage. AlGaN/AlGaN HEMTs also exhibit smaller reductions in cutoff frequency and even higher f\textsubscript{T} values at 300$^\circ$C compared to GaN-channel devices \cite{2012(1)}, a result linked to higher effective electron velocity. However, these thermal benefits come with trade-offs. Alloy disorder and interface roughness in AlGaN reduce mobility, and devices show increased subthreshold swing (SS) above 350°C and virtual gate formation beyond 400$^\circ$C, potentially impacting enhancement-mode reliability. AlGaN channels are more challenging to grow than GaN due to increased strain, composition sensitivity, and surface roughness at high Al content. These challenges may limit the scalability of high-Al-content AlGaN channel HEMTs for high-temperature applications.

In summary, AlGaN-channel HEMTs offer superior high-temperature performance by suppressing leakage (\textbf{Fig. \ref{combined}f}) and improving gate control. Ongoing work aims to develop alternative channel materials or hybrid structures to achieve thermal resilience without sacrificing mobility. \cite{2010I}.

\subsubsection*{Substrate}
A variety of substrates are employed in III-nitride HEMT fabrication, including Si, SiC, sapphire, AlN, and native GaN. These substrates differ significantly in lattice and thermal properties, cost, and scalability. Comparative studies using identical device structures and metal contacts on sapphire, insulating 6H-SiC, semi-insulating 4H-SiC, and Si(111) have shown that all devices experience reduced transconductance ($g_m$) and drain current with increasing temperature, primarily due to phonon-scattering-induced reductions in 2DEG mobility \cite{2002(2),2008(1)}. 

HEMTs on Si and SiC substrates demonstrate lower source and drain resistance at room temperature due to superior thermal conductivity and heat spreading, which supports better performance up to 300$^\circ$C. Above 300$^\circ$C, the drain current is strongly reduced by the mobility degradation, lowering the absolute power dissipation. With less heat dissipated, the channel–ambient temperature rise becomes small; as a result, the influence of substrate thermal conductivity is significantly diminished and the DC characteristics, such as $g_m$ and drain current, are largely substrate independent \cite{2002(2)}. Interestingly, some studies show that Si-based devices may suffer from thermal measurement artifacts at elevated temperatures. Hall measurements on high-resistivity Si (111) at 400$^\circ$C showed apparent sheet charge instability due to thermally generated carriers leaking from substrate to contact \cite{2024K}. In contrast, devices on semi-insulating 4H-SiC exhibited stable sheet charge, underscoring the importance of substrate electrical resistivity and thermal robustness in accurate high-temperature characterization.

Emerging approaches aim to address these limitations by introducing intermediate layers or using doped native substrates. AlGaN/GaN HEMTs grown on a 3C-SiC/high-resistivity Si composite substrate were tested up to 125$^\circ$C, demonstrating that the 3C-SiC layer effectively mitigates lattice and thermal mismatch and suppresses melt-back reactions~\cite{2021(1)}. Separately, AlGaN/GaN HEMTs fabricated on Fe-, C-, and Mn-doped GaN substrates were evaluated up to 327$^\circ$C~\cite{2024(5)}. Fe-doped GaN HEMTs exhibited elevated gate leakage and increased current collapse due to potential distribution and trapping within the doped substrate. C- and Mn-doped GaN substrates significantly reduced leakage, supporting their feasibility for high-temperature GaN HEMTs. Ultimately, the choice of substrate reflects a trade-off between thermal/electrical performance and manufacturability. Although the effects of substrate thermal conductivity can be minimized above 300$^\circ$C, their influence on leakage, reliability, and fabrication scalability remains critical for practical high-temperature HEMT design.


\subsubsection*{Passivation and Dielectric}
Passivation layers are widely used in III-nitride devices to protect the surface from ambient exposure, suppress current collapse by reducing surface trap states, and provide mechanical stability. While performing reliably under room temperature, the passivation layer faced a distinct challenge at high-temperatures. The mechanical and interfacial integrity of passivation can be compromised, requiring careful evaluation to ensure long-term stability.

The effect of passivation layers on 2DEG stability at high temperatures has been extensively studied. Some support the usage of passivation layers, noting improvements, while others observe adverse effects. PECVD SiN$_{x}$ passivation on AlGaN/GaN can effectively suppress strain relaxation and preserve the material integrity after aging up to 900$^\circ$C \cite{2010(2)}. Similarly, PECVD SiN$_{x}$ on AlGaN has enabled HEMT operation for 500 hours at 300$^\circ$C \cite{2008(2)}. Atomic layer deposition (ALD) Al$_{2}$O$_{3}$ passivation on AlGaN sustained HEMT stability up to 747$^\circ$C, with 15\% higher 2DEG density due to surface state reduction or increased polarization \cite{2010}; 10 nm passivation performed the best. 

Despite reported benefits, some studies linked failures or inferior performance to passivation. InAlN/GaN HEMT tested up to 900$^\circ$C for 50 hours with Si$_{3}$N$_{4}$ passivation exhibited no metallization damage, yet cracks in the passivation at the mesa edge were implicated in device failure \cite{2010(3)}. Additionally, the electron mobility and 2DEG density in AlGaN/GaN heterostructures were examined at 600$^\circ$C over 5 hours in both air and argon environments, using van der Pauw with and without ALD Al$_{2}$O$_{3}$ passivation \cite{2017}. After thermal aging, 2DEG density decreased in passivated ones, attributed to partial strain relaxation in the AlGaN layer promoted by combined tensile strain from both Al$_{2}$O$_{3}$ and GaN. Degradation was more pronounced in an argon environment, where no oxidation reaction could heal the cracks in the GaN cap and AlGaN layer or pin the dislocations to suppress strain relaxation. These findings suggest that Al$_{2}$O$_{3}$ passivation may exacerbate strain relaxation at temperature.

This discrepancy largely stems from variations in deposition processes, materials stack configuration, and the thermal stress environment. Among these factors, the nature of stress induced—compressive or tensile—plays a critical role in modulating 2DEG conductivity and overall device stability. For example, high-frequency PECVD Si$_{3}$N$_{4}$ (HF-Si$_{3}$N$_{4}$) on AlGaN has demonstrated superior thermal stability than low-frequency PECVD Si$_{3}$N$_{4}$ (LF-Si$_{3}$N$_{4}$) over 170 hours of operation at 500$^\circ$C \cite{2004}. LF-Si$_{3}$N$_{4}$ introduced compressive stress, reducing the piezoelectric polarization and degrading 2DEG properties. In contrast, HF-Si$_{3}$N$_{4}$ , with a denser film, induced tensile stress, which enhanced piezoelectric polarization and preserved 2DEG conductivity. This stability is linked to the strain solidification in the AlGaN epilayer, driven by the denser Si$_{3}$N$_{4}$ layer. X-ray Diffraction (XRD) confirmed that such dense passivation suppresses strain relaxation in unpassivated AlGaN layers, particularly in thicker layers. 
Meanwhile, the effect of Si$_{3}$N$_{4}$ passivation under thermal stress was studied in Al$_{0.22}$Ga$_{0.78}$N/GaN HEMT with 50nm and 100nm AlGaN  barrier layer, up to 540$^\circ$C \cite{2006}. In passivated 100 nm HEMTs, the in-plane strain increased up to 250$^\circ$C due to thermal expansion mismatch, enhancing piezoelectric polarization and boosting 2DEG concentration. However, the passivated 50 nm HEMTs demonstrated significant strain relaxation above 200$^\circ$C. This relaxation is attributed to the layer’s proximity to the critical thickness, where the increased tensile strain from passivation leads to relaxation through the formation of cracks or dislocation motion once tensile stress exceeds a critical value. 

These studies highlight that passivation is an effective strategy for enhancing high-temperature stability in III-N HEMTs, but its success depends strongly on the material choice, deposition method, and thermal stress compatibility. At elevated temperatures, strain redistribution across the heterostructure can alter polarization and interfacial integrity, making the interplay between layers critical to long-term reliability. As with barrier design, suppressing strain relaxation and preserving interface integrity are key to ensuring stability under thermal stress.
\begin{figure}[h!]
\includegraphics[width=1.025\linewidth]{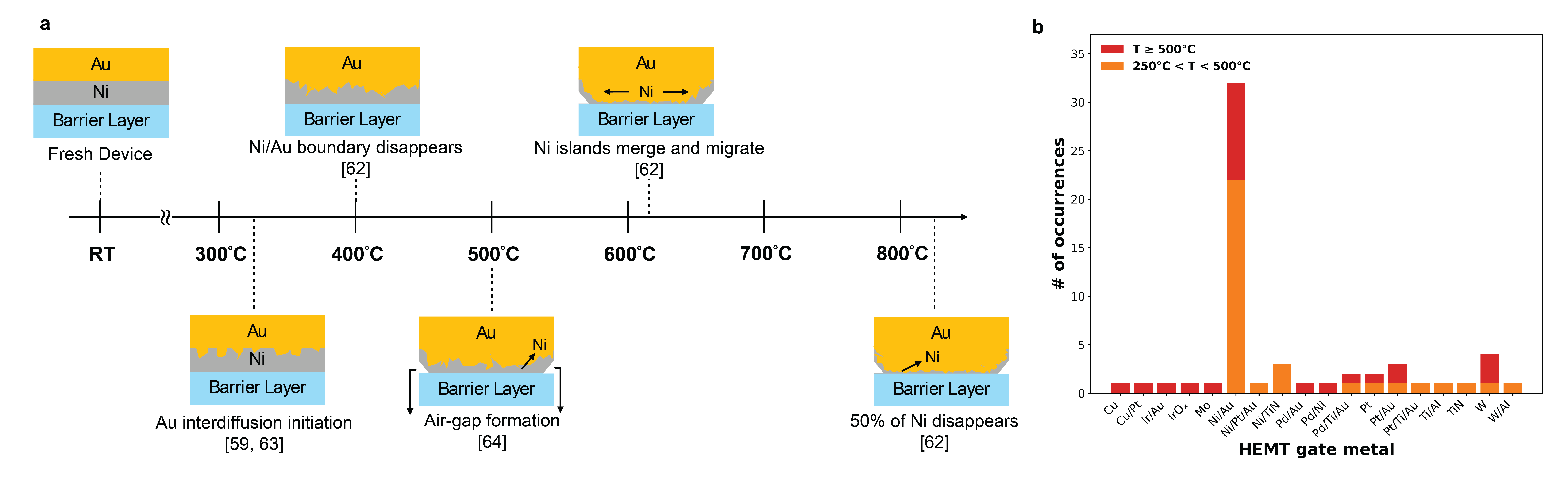}
\caption{\textbf{Thermal degradation pathways of Ni/Au gates and prevalence of gate metal choices in III-N HEMTs
a)} Schematic illustration of the temperature-dependent morphological evolution of conventional Ni/Au gate electrodes on III-N HEMTs.  \textbf{b)} Frequency distribution of gate metals reported in the III-N HEMT literature, categorized by experiments performed at temperatures between 250–500 °C (orange) and $\geq$500 °C (red). Ni/Au appears overwhelmingly as the dominant gate metallization despite reported degradation, while refractory metals such as Pt, Pd, and W appear less frequently.}\cite{Medjdoub2006, 2012(2), 2010(3), 1999, 2005(2), 2009(1), 2024m, Herfurth2013, 2016(1), 2021(6), 2023(7), 2002(1), 2000, 2002(2), 2006(2), 2015F, 2019(1), 2019(3), 2022(5), 2022(6), 2024(1), islam2024effect, 2024K, yuan2023enhancement, 2015(6), 2024(3), 2015(1), 1999e, 2003, 2017(6), 2023(1), 2007(1), 2005(1), 2008(1), 2017(1), cuerdo2009high, 2024(5), 2010(1), 2011(2), 2017(4), 1996, 1997, 2008(2), 2010(5), 2010I, 2012(1), 2012(3), 2014, 2020A, 2023(4), 2006(3), 2011(1), 2018G, 2019B, 2019C, 2021(9)}
\label{metals}
\end{figure}

MISHEMTs introduce new considerations for thermal stability. Comparative studies of ALD Al$_2$O$_3$ MISHEMTs versus Schottky-gated HEMTs up to 600$^\circ$C reveal notable advantages. Unlike Schottky-gated HEMTs exhibiting gate metal diffusion and control degradation, MISHEMTs remained structurally intact at 600$^\circ$C, with reduced leakage attributed to the dielectric serving as an effective diffusion barrier \cite{2016(1)}, resulting in a superior $V_{\text{th}}$ stability. Similarly, Plasma-Enhanced ALD aluminum oxynitride (AlON) has been tested at 500$^\circ$C and exhibited a record high drain current density\cite{2025J}.

Despite these advantages, MISHEMTs can degrade at high temperatures due to crystallization of amorphous dielectrics such as Al$_2$O$_3$ or HfO$_2$, which may cause leakage through grain boundaries \cite{2024O}. Fixed charge and interface traps can shift $V_{\text{th}}$ and degrade subthreshold performance, particularly after thermal cycling, with some MISHEMTs showing increased hysteresis and degraded $g_m$ under extended high-temperature stress \cite{2021epi}. To address these limitations, single-crystal dielectrics such as epitaxial Nd$_2$O$_3$ and Gd$_2$O$_3$ have been explored. Devices incorporating these materials demonstrate stable leakage at 200$^\circ$C, with Gd$_2$O$_3$ showing excellent on/off ratios and minimal $V_{\text{th}}$ shift. Nd$_2$O$_3$-based devices exhibit slightly higher leakage and $V_{\text{th}}$ degradation due to thermally activated interface traps and negative charge trapping. Nonetheless, both materials show promise for improving gate reliability, though longer-duration studies $\geq$300$^\circ$C are needed.

\subsubsection*{Device Geometry and Scaling}
The geometry of the devices plays an essential role in their behavior across various conditions, including high temperatures. Up to 500$^\circ$C, the reduction in saturation current is more severe in longer gate length devices \cite{2006(2), 2008(1)}. Shorter gate lengths maintain higher electric field, placing carriers into the velocity saturation regime where carrier velocity - and thus current - is less sensitive to temperature, with its dependence relaxing from \(T^{-1.5} \, \text{to} \, T^{-0.5}\). 
Circular HEMTs (C-HEMT) have demonstrated superior thermal resilience compared to conventional linear HEMTs (L-HEMT) at elevated temperatures \cite{2017inaln, 2025investigation}. L-HEMTs, which use mesa-defined active regions, are vulnerable to traps generated at etched sidewalls. In contrast, C-HEMTs feature a concentric layout with the gate electrode surrounding the drain, eliminating the need for mesa isolation and enabling more uniform current control. As a result, C-HEMTs remained functional up to 600$^\circ$C, whereas L-HEMTs suffered permanent device failure. The enclosed geometry of C-HEMTs is likely to suppress sidewall-induced traps and reduce leakage at high temperatures. The off-state current of the L-HEMT was three orders of magnitude larger than that of the C-HEMT \cite{2023(1)}, driven by the two-dimensional variable range hopping (2D-VRH) conduction along etched GaN surface, which dominates up to 300$^\circ$C. This conduction mechanism is not observed in C-HEMTs, supporting their thermal robustness. While the circular layout offers a promising approach to suppress thermally activated leakage, it is not practical for large-area industrial applications due to layout inefficiency and packaging constraints, but has been utilized for pressure-sensing at high-temperature \cite{2017inaln}. In contrast, ion-implanted isolation, which is widely adopted in industry, remains underexplored in terms of its high-temperature behavior and warrants further investigation.

\subsubsection*{Contacts}
Metal contact degradation is a major concern in the high-temperature operation of III-nitride HEMTs. When device operating temperatures approach or exceed these formation conditions, further interdiffusion, alloying, and Au penetration into underlying layers can occur, leading to contact instability and increased resistivity. Ohmic stacks are conventionally composed of Ti/Al/X/Au, where \( X \) acts as a diffusion barrier layer to Au in-diffusion into the Ti/Al region and GaN interface. Refractory metals such as Ni, Mo, Ta, Ir, Nb, and Pt are commonly used due to their superior thermal stability \cite{2016}. Ti/Al/Ni/Au contacts to AlGaN/GaN heterostructures have demonstrated stability after 48 hours of thermal stress at 325$^\circ$C, while Ti/Al/Mo/Au contacts maintained consistent performance over 244 hours at the same temperature \cite{2017(4), 2021(6)}. Similarly, at 500$^\circ$C, both Ti/Al/Ni/Au and regrown \( n^{++} \)GaN Ohmic contacts maintained functional stability for 24 hours, but exhibited rapid increases in sheet resistance and surface roughening after 48 hours, correlated with AlGaN barrier degradation and disruption of the 2DEG/regrown GaN interface. Transmission Electron Microscopy (TEM) and Energy Dispersive Spectroscopy (EDS) confirmed that the bulk contact structure remained intact despite interfacial changes \cite{2024J}. However, in scaled-length devices, where contact resistance dominates, increased contact resistivity – driven by interfacial degradation, metal migration, and enhanced phonon or defect scattering – becomes a critical limiter of current drive and $g_m$. Moreover, in situ TEM studies of AlGaN/GaN HEMTs with Ti/Al/Ni/Au contacts reveal Ti/GaN intermixing and contact discontinuities at 800$^\circ$C due to metal evaporation and migration \cite{2024m}. Despite these challenges, Ohmic contacts are generally not considered the dominant failure mechanism in high-temperature GaN HEMTs, though careful contact stack design and control of Au diffusion remain essential for extending operational lifetimes.

Unlike Ohmic contacts, Schottky gate contacts in III-nitride HEMTs are highly susceptible to thermal degradation. Ni/Au, the most widely used gate metal stack, has been repeatedly shown to degrade structurally and electrically when exposed to elevated temperatures starting at 325$^\circ$C. Scanning electron microscopy has revealed early-stage intermixing and void formation in Ni/Au gates at 325$^\circ$C \cite{Macron2011, 2017(4)}, leading to increased gate and drain leakage. Gate-defining etches exacerbate this by introducing voids and sidewall damage, promoting localized failures. By 400$^\circ$C, distinct interfaces disappear, indicating alloying between Ni and Au \cite{2024m}, and by 500$^\circ$C, the formation of air gaps near the gate region and loss of the Ni layer further reduce effective work function and increase gate leakage \cite{islam2024effect}. At 600$^\circ$C, Ni segregates into metal islands that migrate laterally toward the gate edge, and by 800$^\circ$C, half the gate metal can be lost through recrystallization, densification, or evaporation \cite{2024m}. These thermal degradation pathways are comprehensively illustrated in \textbf{Fig. \ref{metals}a}. 

Although widely used due to fabrication ease and legacy compatibility (\textbf{Fig. \ref{metals}b}), Ni/Au's limitations highlight the need for more thermally stable alternatives. Ni/Pt/Au gates on AlGaN/GaN HEMTs degraded after 1000 hours at 350$^\circ$C due to Au diffusion circumventing the Pt barrier, leading to a two-order-of-magnitude increase in gate leakage \cite{2017(1)}. Similarly, Pt/Ti/Au gates exhibit discoloration above 300$^\circ$C and alloying by 400$^\circ$C\cite{1996}. IrO$_x$ gates on InAlN/GaN HEMTs remain intact at 600$^\circ$C in air, but suffer irreversible degradation upon cooling \cite{2021(6)}. Mo/Au gates exhibit a ~50\% reduction in drain current and 70\% decrease in on/off ratio after 120 hours of stress at 472$^\circ$C and 465$^\circ$C in air and simulated Venus conditions, respectively \cite{2010(3), 2025investigation}. Cu gates on InAlN/GaN HEMTs withstood 900$^\circ$C in vacuum without visible metallization breakup; however, Cu electromigration and passivation rupture were noted \cite{2010(3)}. 

Pt/Au gates remain stable at 240$^\circ$C for 48 hours \cite{2004}, while Ir/Au gates show stability for 120 hours at 500$^\circ$C \cite{2000}. Tungsten (W) gates on enhancement-mode p-Gan/AlGaN/GaN HEMTs are shown to be structurally robust after 10 days in a Venus-like environment \cite{yuan2023enhancement}. Similarly, Pd/TaSi$_2$ on p-Gan/AlGaN/GaN has been tested up to 800$^\circ$C \cite{2025pGaN800}, sustaining record-high on/off ratios at 800$^\circ$C without noticeable degradation during one hour of operation. Beyond these demonstration, a broader set of gate materials—including Cu/Pt \cite{Herfurth2013}, Pd/Au \cite{2019(3)}, Pd/Ni \cite{2015F}, Pd/Ti/Au \cite{2002(2)}, Ti/Al \cite{2023(4)}, Ti/Al/TiN \cite{2021(8)}, Ni/TiN \cite{2019B}, and TiN \cite{2023(4)}—have been preliminarily explored at high temperatures (\textbf{Fig. \ref{metals}b}). Systematic long‑term evaluation for these materials is still required to fully assess their reliability.

\subsection*{Device Physics}
To better understand the thermal limitations of HEMTs, it is essential to become familiar with the mechanisms behind their performance degradation.

The primary scattering processes in III-N-based HEMTs are polar optical phonon (POP) scattering, acoustic phonon scattering, ionized impurity scattering, interface roughness scattering, and alloy scattering \cite{wide_bandgap}. Only POP and acoustic phonon scattering exhibit a strong temperature dependence. Acoustic phonon scattering dominates at temperatures between 10 to 150K, whereas POP scattering dominates above 150 K \cite{2010, 1996bulk}. Due to GaN's polar nature \cite{2006(2),1999}, optical phonons strongly interact with 2DEG electrons, reducing mobility as \(T^{-1.8} \, \text{to} \, T^{-2.6}\). Beyond 400$^\circ$C, mobility degradation increases sheet resistance, and reduces drain current and $g_m$, making AlGaN/GaN comparable to InAlN/GaN\cite{2024K, 2016(1)}.To mitigate POP scattering in HEMTs, strategies such as material engineering (alloying, heterostructures), interface optimization, carrier screening, and thermal management may be pursued. Modifying electron-phonon coupling can bolster mobility at high temperatures.

$V_{\text{th}}$ (\textbf{Fig. \ref{combined}g}) shifts with temperature in both E-mode and D-mode HEMTs. Generally in E-mode HEMTs, V\textsubscript{th} decreases (negative shift) with rising temperature\cite{2007(1), 2023(8), 2015(1), 2023(1), 2024(1)}, whereas it typically increases (positive shift) in D-mode\cite{2007(1), 2016(1), 2022(5), 2017(4)}.Degradation is driven by strain relaxation and activation of interface traps \cite{2016(1)}. Strategies such as recessed-barrier E-mode MISHEMTs have shown enhanced $V_{\text{th}}$ stability up to 400$^\circ$C \cite{2023(8)} due to reduced trap activation at the interface. In contrast, D-mode MISHEMTs suffer from deeper interface traps between 350-400$^\circ$C, leading to $V_{\text{th}}$ instability\cite{2023(1)}.

As in Si MOSFETs, the interface trap density plays a dominant role in high-temperature SS deterioration in GaN HEMTs \cite{2007_for_SS}. Rising temperature activates traps, increasing SS and reducing efficiency \cite{ 2019(1), 2023(1), 2024(1)}. To address this, ALD-deposited Al$_2$O$_3$ has been used as an etch-stop layer in p-GaN/AlGaN/GaN HEMTs operating up to 500$^\circ$C \cite{2024(1)}. This effectively suppresses hole injection at the p-GaN/Schottky metal interface, enhancing gate control and minimizing hysteresis, resulting in a lower SS.


Pulse measurements assess high-temperature trapping. At 300$^\circ$C, drain lag observed in AlGaN/GaN HEMTs was linked to current collapse from surface trapping \cite{2008(2)}. AlGaN/GaN HEMTs show stable $V_{\text{th}}$ for 192 hours at 325$^\circ$C, aided by SiN$_{x}$ passivation \cite{2010(1)}. Beyond 384 hours, interdiffusion-induced leakage indicate gate contact, not trap, failure \cite{Macron2011}. At higher temperatures (440$^\circ$C), drain–gate leakage escalates\cite{2015(6)}. At 500$^\circ$C, both gate and drain pulse measurements reveal increased gate leakage and drain current reduction due to trapping in the barrier, inducing a virtual gate not observed at room temperature. \cite{2019(1), islam2024effect}.

Future efforts should aim to minimize trap activation and stabilize $V_{\text{th}}$ and SS above 500°C. Such shifts impair switching efficiency and increase leakage, degrading III-nitride circuit performance in high-temperature logic, power, and RF systems. 

\subsection*{High-Temperature Integrated Circuits and Systems}
System-level evaluation is essential to understanding III-nitride HEMT performance at high temperatures. Inverters are a logical starting point for evaluating circuit-level behavior. Multiple studies have reported the thermal performance of GaN HEMT and MIS-HEMT direct-coupled FET logic (DCFL) inverters (\textbf{Fig. \ref{failure-circuits}}). In a comparison between a p-GaN-gated E-mode/D-mode (E: driver, D: load) and E-mode/E-mode HEMT inverter,  E/D-mode performed better at 300$^\circ$C in terms of voltage swing, gain, and noise margin ($NM$) \cite{2022(6)}. These results highlight the importance of integrating E-mode and D-mode GaN HEMTs. E/D-mode swing scales better with $V_{DD}$ due to constant D-mode load current.


Moreover, an E/D-mode HEMT inverter has been successfully demonstrated to operate at 500$^\circ$C. Its voltage transfer characteristics (VTC) remained consistent up to 400$^\circ$C, with only slight decrease in $NM_L$ due to the degradation of the on-resistance in the E-mode driver as compared to the D-mode load.\cite{2022(6)} Another work on E/D-mode AlGaN/GaN HEMT inverters has similarly reported stable operation at 375$^\circ$C. Undesirable performance shifts were caused by negative $V_{\text{th}}$ shift of the E-mode driver with increasing temperature. It was proposed that the noise margin could be enhanced by raising the E-mode $V_{th}$ \cite{2007(1)}. In addition to HEMT/HEMT configurations, E-mode HEMTs have also been integrated with D-mode MISHEMTs to create inverters operational at 300$^\circ$C\cite{2023(4)} and 500$^\circ$C \cite{2024(1)}. GaN MIS-HEMT/MIS-HEMT studies have been conducted at 400$^\circ$C \cite{2023(1)} and 250$^\circ$C, likewise noting a left-shifted VTC curve.\cite{2019C} As single III-nitride HEMTs can operate up to 1000$^\circ$C,\cite{2012(2)}, there is still much opportunity to push the high temperature performance of DCFL inverters.

III-nitride HEMTs have also been incorporated into more complex digital blocks such as ring oscillators,\cite{2022(6), 2007(1)} combinational logic,\cite{2023(4), 2023Towards} and memory cells.\cite{2022gan} Ring oscillators measure propagation delay, $\sim$0.1 ns/stage at RT, which worsens at high temperature due to reduced drive current\cite{2022(6), 2007(1)}. When testing AlGaN/GaN HEMT NAND and NOR gates, channel electron mobility and on-current again fell with temperature causing slower transitions, but correct logic output persisted to 300$^\circ$C \cite{2023(4)}. A functional GaN ROM, SRAM, D flip-flop, D latch,\cite{2022gan} ALU\cite{2023Towards} have all been proven operational up to 300$^\circ$C, indicating enormous potential in high-temperature circuitry. Aggressively scaling down transistor size in order to boost the current density with respect to gate capacitance has been suggested \cite{2022gan}.

Yet, E/D-mode GaN circuits suffer from high static power and limited swing compared to CMOS technology, motivating interest in combined n/p-channel GaN-based logic (\textbf{Fig. \ref{failure-circuits}}). p-channel III-nitride HEMTs rely on a two-dimensional hole gas (2DHG). The high-temperature operation of p-GaN HEMTs has been studied up to 400$^\circ$C without degradation \cite{2024pGaN}. Complementary GaN HEMTs have been operated at 350$^\circ$C \cite{2024HOLE2}, and CMOS-like GaN HEMT inverter has operated at 300$^\circ$C.\cite{2020A} However, the significant limiting factor in attaining high-efficiency CMOS-like III-nitride HEMT circuits is the poor hole mobility, which remains an issue to be addressed\cite{2016CMOS}.

Some studies address how temperature impacts RF performance in III-nitride HEMTs. Experimental data on microwave parameters, including cut-off frequency (f$_{T}$) and maximum oscillation frequency (f$_{max}$), have been recorded up to 500$^\circ$C \cite{2021(1), cuerdo2009high, ottaviani2020evaluation, akita2001high, xue2024measurement, cui2022scaling}. In all studies, it has been consistently shown that f$_{T}$ and f$_{max}$ monotonically decrease with rising temperature (\textbf{Fig. \ref{combined}h}). This reduction is primarily attributed to a decrease in effective electron velocity, stemming from reduced carrier mobility at high temperatures. Simultaneously, intrinsic capacitances C$_{gs}$ and C$_{gd}$ exhibit a moderate increase due to temperature-induced changes in the permittivity of both the barrier layer and the passivation layer \cite{cuerdo2009high, akita2001high}. More high-T RF data above 250$^\circ$C is needed. For instance, although GaN HEMT low-noise amplifiers (LNAs) and power amplifiers (PAs) have been demonstrated utilizing III-nitride HEMTs, but few studies are reported at high temperature and multi-stage GaN HEMT LNA/PA operation and systematic RF characterization (e.g., noise figure, gain compression (P1dB), linearity (IIP3)) $\geq$250$^\circ$C remains unexplored (\textbf{Fig. \ref{failure-circuits}}). A GaN bootstrapping amplifier was recently operated for 60 min at 800$^\circ$C, exemplifying the promise of high temperature GaN HEMT technology \cite{2025bootstrap}.


HEMT power conversion circuits above 200$^\circ$C are rare - only a GaN-based DC-DC buck converter at 250$^\circ$C \cite{2019monolithic} and a flyback converter at 200$^\circ$C \cite{2016highddd} have been reported. Field plates (FP) (\textbf{Fig. \ref{failure-circuits}}), commonly employed in HEMTs at room temperature to improve field distribution and breakdown voltage, represent an open area to be explored for high‑temperature operation, as neither the fundamental breakdown limits set by material properties nor the effectiveness of field‑plate engineering have been thoroughly investigated under elevated‑temperature conditions \cite{2022optimization}. Overall, GaN HEMTs show strong potential across logic, analog, and power domains with numerous advancements and opportunities for advancement in high-temperature deployment waiting to be realized.

\section*{Outlook} 
The role of III-nitride HEMTs in extreme environment electronics is expanding as they consistently demonstrate operation at temperatures above the limits of silicon. However, there remains a persistent gap between short-term demonstrations and long-term, high-reliability performance needed for the most demanding applications. Closing this gap will require coordinated progress in materials engineering, device architecture, circuit integration, and environmental testing. At the materials level, several strategies have emerged to address the dominant thermal failure mechanisms, such as AlGaN channels and ScAlN barriers. Replacing conventional Ni/Au gate contacts with refractory metals such as molybdenum, iridium \cite{newaddedIrOx}, or tungsten also shows promise for improving $V_{\text{th}}$ stability.

Due to their fabrication simplicity and high current capability, depletion-mode devices have been more extensively studied than enhancement-mode devices, but the latter are preferable for applications requiring low standby power, fail-safe behavior, and logic compatibility. However, enhancement-mode devices are also more sensitive to $V_{\text{th}}$ shifts and interface degradation under thermal stress. Therefore, future research should include side-by-side evaluations of these architectures under long-term bias and temperature exposure, especially in circuit-relevant configurations.

Another critical consideration is the testing environment. Many reported high-temperature studies are conducted under vacuum or inert ambient conditions, which do not capture the effects of chemically active or reactive environments in real applications. For example, hypersonic vehicles operate in high-pressure oxygen-rich air, and planetary probes, geothermal, and nuclear systems may be exposed to corrosive gases and radiation. These environments can accelerate HEMT degradation. Realistic testing that includes controlled exposure to air, vacuum, carbon dioxide, or mixed gases is essential for understanding the true reliability of III-nitride HEMTs in these environments. Although SiC remains the established benchmark for high-voltage and long-duration high-temperature electronics, its doped bulk conduction and gate-oxide limitations contrast with the polarization-driven, heterostructure-based design of GaN \cite{sIc2024review}. We suggest that SiC and GaN occupy complementary performance regimes rather than a hierarchy, each suitable for different domains of the high-temperature electronics landscape.

There is an urgent need for compact models that accurately capture thermal effects, including mobility degradation, trap dynamics, gate leakage, and $V_{\text{th}}$ shift. Existing models are typically calibrated for near-room-temperature operation and fail to reflect performance in high-temperature or chemically reactive environments. To support reliable circuit design, these models must be extended and validated under conditions representative of real-world applications. Equally important, future progress will require tight co-optimization between HEMT device structures and circuit architectures: material-level degradation mechanisms directly influence circuit metrics like timing margin, voltage swing, and power efficiency, while circuit loading and biasing conditions affect device stress and reliability. Bridging this gap will be essential for the development of GaN-based logic, radio frequency, and power systems capable of robust operation above 300$^\circ$C. Although this review has focused on lateral Ga-polar HEMTs due to their process maturity, emerging N-polar and vertical GaN devices offer interesting opportunities. Although they are still in the early stages of development, they may be better suited for applications that require compact footprints, high power density, and enhanced thermal management.

The most promising route to stable high-temperature III-N HEMTs likely involves combining lattice-matched InAlN or AlGaN channel/barrier stacks with refractory gate metals (such as Ir, W, or Pd/TaSi$_2$) and MIS-HEMT designs that incorporate thermally robust dielectrics like epitaxial Gd$_2$O$_3$. These materials systems offer strong temperature resilience and, if integrated into enhancement-mode HEMTs, could enable both high reliability and logic compatibility. Additional gains may come from circular or ion-implanted isolation and high-frequency PECVD passivation, which help to mitigate leakage and suppress strain relaxation. However, the main roadblock remains the lack of long-term reliability data under chemically reactive, high-temperature environments. Testing in vacuum or inert ambient does not capture real-world degradation from oxidizing, corrosive, or radiative conditions. Without standardized lifetime qualification under such stresses, practical deployment will remain limited. Addressing this will require improved test platforms, predictive models, and closer links between device reliability and circuit-level performance, as well as a deeper understanding of the critical interdependencies among gate stack integrity, thermal conduction pathways, and defect dynamics that shape system-level behavior. Quantifying degradation mechanisms across a range of conditions, including device scaling, circuit load, and realistic ambient exposure, will be critical to unlocking the full device potential. With sustained investment and coordinated research across disciplines, III-N HEMTs can become a foundational technology for electronics that must operate reliably in the most extreme conditions.

\bibliography{sample}

\noindent\textbf{Acknowledgments}\\
This work was sponsored in part by the Air Force Office of Scientific Research (AFOSR) under award no. FA9550-22-1-0367 and Lockheed Martin Corporation under award no. 025570-00036.

\noindent\textbf{Author contributions}\\
Y.L., J.Z., and J.N. wrote the main text and prepared the figures. H.P. collected data. T.P. provided supervision, technical assistance, and co-wrote the outlook section. S.R.E. developed the initial outline, edited the manuscript for form and content, and co-wrote the outlook section. 

\noindent\textbf{Competing interests}\\
The authors declare no competing interests.


\end{document}

%% file: table_barrier_props.tex
\begin{table}[ht!]
\small
\centering
\caption{Fundamental Material Properties of Common III-N Barrier Materials for GaN HEMTs}
\begin{tabularx}{\textwidth}{>{\bfseries}l X X X X}
\toprule
Property & AlN & Al$_x$Ga$_{1-x}$N & In$_x$Al$_{1-x}$N & Sc$_x$Al$_{1-x}$N \\
\midrule
\textbf{Typical Thickness} 
& 2–3\,nm 
& 15–25\,nm 
& 8–12\,nm 
& 5–10\,nm \\

\textbf{Typical $x$ Content} 
& $x = 1$ 
& $x \approx 0.2–0.3$ 
& $x \approx 0.17–0.18$ 
& $x \approx 0.2–0.3$ \\

Mobility~\cite{2024K, lee2015temperature, Alpert2020SensitivityTemperatures,lee2015temperature, islam2024effect, JohnThesis} 
& High (900–1700 cm$^2$/V$\cdot$s) 
& High (900–2200 cm$^2$/V$\cdot$s) 
& Moderate (600–1400 cm$^2$/V$\cdot$s) 
& Lower (300–800 cm$^2$/V$\cdot$s) \\

2DEG Carrier Concentration~\cite{2024K, lee2015temperature, Alpert2020SensitivityTemperatures} 
& $\sim$0.7–3.1 $\times 10^{13}$ cm$^{-2}$ 
& $\sim$0.7–1.5 $\times$ 10$^{13}$ cm$^{-2}$ 
& $\sim$1.5–3 $\times$ 10$^{13}$ cm$^{-2}$ 
& $>$2 $\times$ 10$^{13}$ cm$^{-2}$ predicted \\

Strain / Lattice Matching~\cite{2024Simulation,Alpert2020SensitivityTemperatures, 2015F} 
& High tensile mismatch with GaN ($\sim$2.4\%) 
& Tensile unless buffered 
& Lattice-matched at $\sim$18\% In 
& Tensile; mismatch with GaN \\

Ease of Growth 
& Very challenging; ultrathin (2–3\,nm) 
& Mature via MOCVD 
& Challenging; In desorption 
& Difficult; grown via MBE/sputtering \\

Technology Maturity 
& Low to moderate (research + niche commercial, e.g., HRL) 
& High (commercial standard) 
& Moderate (academic + early industrial) 
& Low (research-stage) \\

High-$T$ Operation~\cite{2012(2),2024ScAlN, 1999} 
& Limited; prone to relaxation and leakage 
& Good ($\sim$400–500$^\circ$C with passivation) 
& Excellent (1000$^\circ$C) 
& Unproven, but promising\\

\textbf{Primary High-$T$ Operation Challenge} 
& Strain relaxation, gate leakage 
& Trapping, thermal mismatch 
& Indium volatility, surface roughness 
& Phase instability, strain relaxation \\
\bottomrule
\end{tabularx}
\label{tab:barrier_props}
\end{table}

%% file: sample.bib
@ARTICLE{1997,
   author = {Gaska, R and Chen, Q and Yang, J and Osinsky, A and Khan, M Asif and Shur, Michael S},
   title = {{High-temperature performance of AlGaN/GaN HFETs on SiC substrates}},
   journal = {IEEE Electron Device Letters},
   volume = {18},
   pages = {492--494},
   year = {1997}
}

@ARTICLE{2024K,
   author = {Xie, Qingyun and Niroula, John and Rajput, Nitul S and Yuan, Mengyang and Luo, Shisong and Fu, Kai and Isamotu, Mohamed Fadil and Palash, Rafid Hassan and Sikder, Bejoy and Eisner, Savannah R and others},
   title = {{Device and material investigations of GaN enhancement-mode transistors for Venus and harsh environments}},
   journal = {Applied Physics Letters},
   volume = {124},
   year = {2024}
}

@ARTICLE{1999,
   author = {Daumiller, I and Kirchner, C and Kamp, M and Ebeling, Karl Joachim and Kohn, E},
   title = {{Evaluation of the temperature stability of AlGaN/GaN heterostructure FETs}},
   journal = {IEEE Electron Device Letters},
   volume = {20},
   pages = {448--450},
   year = {1999}
}

@ARTICLE{2006,
   author = {Chen, DJ and Shen, B and Zhang, KX and Tao, YQ and Wu, XS and Xu, J and Zhang, R and Zheng, YD},
   title = {{Temperature dependence of strain in Al0. 22Ga0. 78N/GaN heterostructures with and without Si3N4 passivation}},
   journal = {Thin solid films},
   volume = {515},
   pages = {4384--4386},
   year = {2007}
}

@ARTICLE{2004,
   author = {Feng, ZH and Zhou, Yu Gang and Cai, SJ and Lau, Kei May},
   title = {{Enhanced thermal stability of the two-dimensional electron gas in GaN/ AlGaN/ GaN heterostructures by Si3N4 surface-passivation-induced strain solidification}},
   journal = {Applied physics letters},
   volume = {85},
   pages = {5248--5250},
   year = {2004}
}

@ARTICLE{2010,
   author = {Tokuda, H and Yamazaki, J and Kuzuhara, M},
   title = {{High temperature electron transport properties in AlGaN/GaN heterostructures}},
   journal = {Journal of Applied Physics},
   volume = {108},
   year = {2010}
}

@ARTICLE{2017,
   author = {Hou, Minmin and Jain, Sambhav R and So, Hongyun and Heuser, Thomas A and Xu, Xiaoqing and Suria, Ateeq J and Senesky, Debbie G},
   title = {{Degradation of 2DEG transport properties in GaN-capped AlGaN/GaN heterostructures at 600° C in oxidizing and inert environments}},
   journal = {Journal of Applied Physics},
   volume = {122},
   year = {2017}
}

@ARTICLE{2016,
   author = {Greco, Giuseppe and Iucolano, Ferdinando and Roccaforte, Fabrizio},
   title = {{Ohmic contacts to Gallium Nitride materials}},
   journal = {Applied Surface Science},
   volume = {383},
   pages = {324--345},
   year = {2016}
}

@ARTICLE{2024J,
   author = {Niroula, John and Xie, Qingyun and Rajput, Nitul S and Darmawi-Iskandar, Patrick K and Rahman, Sheikh Ifatur and Luo, Shisong and Palash, Rafid Hassan and Sikder, Bejoy and Yuan, Mengyang and Yadav, Pradyot and others},
   title = {{High temperature stability of regrown and alloyed Ohmic contacts to AlGaN/GaN heterostructure up to 500° C}},
   journal = {Applied Physics Letters},
   volume = {124},
   year = {2024}
}

@ARTICLE{2024m,
   author = {Rasel, Md Abu Jafar and Zhang, Di and Chen, Aiping and Thomas, Melonie and House, Stephen D and Kuo, Winson and Watt, John and Islam, Ahmad and Glavin, Nicholas and Smyth, M and others},
   title = {{Temperature-induced degradation of GaN HEMT: An in situ heating study}},
   journal = {Journal of Vacuum Science \& Technology B},
   volume = {42},
   year = {2024}
}

@article{islam2024effect,
  author={Islam, Ahmad E and Sepelak, Nicholas P and Miesle, Adam T and Lee, Hanwool and Snure, Michael and Nikodemski, Stefan and Walker, Dennis E and Miller, Nicholas C and Grupen, Matt and Leedy, Kevin D and others},
  title={{Effect of High Temperature on the Performance of AlGaN/GaN T-Gate High-Electron Mobility Transistors With~ 140-nm Gate Length}},
  journal={IEEE Transactions on Electron Devices},
  volume={71},
  number={3},
  pages={1805--1811},
  year={2024},
}

@ARTICLE{1996,
   author = {Aktas, Ozgur and Fan, ZF and Mohammad, SN and Botchkarev, AE and Morkoc, H},
   title = {{High temperature characteristics of AlGaN/GaN modulation doped field-effect transistors}},
   journal = {Applied Physics Letters},
   volume = {69},
   pages = {3872--3874},
   year = {1996}
}

@ARTICLE{2000,
   author = {Hilsenbeck, J and Nebauer, E and Wurfl, J and Trankle, G and Obloh, H},
   title = {{Aging behaviour of AlGaN/GaN HFETs with advanced ohmic and Schottky contacts}},
   journal = {Electronics Letters},
   volume = {36},
   pages = {1},
   year = {2000}
}

@ARTICLE{2015F,
   author = {Gaska, Remis and Gaevski, M and Jain, R and Deng, J and Islam, M and Simin, G and Shur, M},
   title = {{Novel AlInN/GaN integrated circuits operating up to 500 C}},
   journal = {Solid-State Electronics},
   volume = {113},
   pages = {22--27},
   year = {2015}
}

@ARTICLE{2003,
   author = {Arulkumaran, S and Egawa, Takashi and Ishikawa, H and Jimbo, Takashi},
   title = {{Temperature dependence of gate--leakage current in AlGaN/GaN high-electron-mobility transistors}},
   journal = {Applied physics letters},
   volume = {82},
   pages = {3110--3112},
   year = {2003}
}

@ARTICLE{2019B,
   author = {Cui, Miao and Bu, Qinglei and Cai, Yutao and Sun, Ruize and Liu, Wen and Wen, Huiqing and Lam, Sang and Liang, Yung C and Mitrovic, Ivona Z and Taylor, Stephen and others},
   title = {{Monolithic integration design of GaN-based power chip including gate driver for high-temperature DC--DC converters}},
   journal = {Japanese Journal of Applied Physics},
   volume = {58},
   pages = {056505},
   year = {2019}
}

@inproceedings{2019C,
  author={Cui, Miao and Cai, Yutao and Bu, Qinglei and Liu, Wen and Wen, Huiqing and Mitrovic, Ivona Z and Talyor, Stephen and Chalker, Paul R and Zhao, Cezhou},
  title={{The impact of etch depth of D-mode AlGaN/GaN MIS-HEMTs on DC and Ac characteristics of 10 V input direct-coupled FET logic (DCFL) inverters}},
  booktitle={2019 International Conference on IC Design and Technology (ICICDT)},
  pages={1--4},
  year={2019},
  organization={IEEE}
}

@ARTICLE{2010I,
   author = {Tokuda, Hirokuni and Hatano, Maiko and Yafune, Norimasa and Hashimoto, Shin and Akita, Katsushi and Yamamoto, Yoshiyuki and Kuzuhara, Masaaki},
   title = {{High Al composition AlGaN-channel high-electron-mobility transistor on AlN substrate}},
   journal = {Applied physics express},
   volume = {3},
   pages = {121003},
   year = {2010}
}

@ARTICLE{2014,
   author = {Yafune, Norimasa and Hashimoto, Shin and Akita, Katsushi and Yamamoto, Yoshiyuki and Tokuda, Hirokuni and Kuzuhara, Masaaki},
   title = {{AlN/AlGaN HEMTs on AlN substrate for stable high-temperature operation}},
   journal = {Electronics letters},
   volume = {50},
   pages = {211--212},
   year = {2014}
}

@ARTICLE{wide_bandgap,
   author = {Takahashi, Kiyoshi and Yoshikawa, Akihiko and Sandhu, Adarsh},
   title = {{Wide bandgap semiconductors}},
   journal = {Verlag Berlin Heidelberg},
   year = {2007}
}

@ARTICLE{1996bulk,
   author = {Shur, M and Gelmont, B and Asif Khan, M},
   title = {{Electron mobility in two-dimensional electron gas in AIGaN/GaN heterostructures and in bulk GaN}},
   journal = {Journal of electronic materials},
   volume = {25},
   pages = {777--785},
   year = {1996}
}

@ARTICLE{2024Simulation,
   author = {Hidayat, Wagma and Usman, Muhammad},
   title = {{Exploring the impact of AlGaN barrier thickness and temperature on normally-on GaN HEMT performance}},
   journal = {Engineering Research Express},
   volume = {6},
   pages = {025307},
   year = {2024}
}

@inproceedings{2024O,
  author={Singh, Umang and Genath, Hannah and Sarkar, Ritam and Kruegener, Jan and Osten, H Joerg and Laha, Apurba},
  title={{Improved thermal stability at high temperature of operation (473 K) in all epitaxy Nd 2 O 3/AlGaN/GaN MOSHEMT}},
  booktitle={2024 8th IEEE Electron Devices Technology \& Manufacturing Conference (EDTM)},
  pages={1--3},
  year={2024},
}

@inproceedings{2024HOLE2,
  author={Luo, Shisong and Chang, Cheng and Xie, Qingyun and Li, Tao and Xu, Mingfei and He, Ziyi and Palacios, Tom{\'a}s and Zhao, Yuji},
  title={{GaN E-mode Complementary Transistors Based on a GaN-on-Si Platform Operational at 350° C}},
  booktitle={2024 Device Research Conference (DRC)},
  pages={1--2},
  year={2024},
}

@ARTICLE{2024p,
   author = {Niroula, John and Taylor, Matthew A and Xie, Qingyun and Yadav, Pradyot and Luo, Shisong and Zhao, Yuji and Palacios, Tom{\'a}s},
   title = {{Record High Temperature Performance in Scaled AlGaN/GaN-on-Si HEMTs up to 500° C}},
   journal = {2024 Device Research Conference (DRC)},
   pages = {1--2},
   year = {2024}
}

@inproceedings{2022Y,
  author={Yuan, Mengyang and Xie, Qingyun and Niroula, John and Isamotu, Mohamed Fadil and Rajput, Nitul S and Chowdhury, Nadim and Palacios, Tom{\'a}s},
  title={{High temperature robustness of enhancement-mode p-GaN-gated AlGaN/GaN HEMT technology}},
  booktitle={2022 IEEE 9th Workshop on Wide Bandgap Power Devices \& Applications (WiPDA)},
  pages={40--44},
  year={2022},
}

@ARTICLE{1999e,
   author = {Maeda, Narihiko and Saitoh, Tadashi and Tsubaki, Kotaro and Nishida, Toshio and Kobayashi, Naoki},
   title = {{Superior pinch-off characteristics at 400 C in AlGaN/GaN heterostructure field effect transistors}},
   journal = {Japanese journal of applied physics},
   volume = {38},
   pages = {L987},
   year = {1999}
}

@ARTICLE{2018G,
   author = {Sokolovskij, Robert and Zhang, Jian and Iervolino, Elina and Zhao, Changhui and Santagata, Fabio and Wang, Fei and Yu, Hongyu and Sarro, Pasqualina M and Zhang, Guo Qi},
   title = {{Hydrogen sulfide detection properties of Pt-gated AlGaN/GaN HEMT-sensor}},
   journal = {Sensors and Actuators B: Chemical},
   volume = {274},
   pages = {636--644},
   year = {2018}
}

@INPROCEEDINGS{2007_for_ss,
  author={Chung, J. W. and Zhao, X. and Palacios, T.},
  title={{Estimation of Trap Density in AlGaN/GaN HEMTs from Subthreshold Slope Study}}, 
  booktitle={2007 65th Annual Device Research Conference}, 
  volume={},
  number={},
  pages={111-112},
  year={2007},
}

@inproceedings{2023Towards,
  author={Xie, Qingyun and Yuan, Mengyang and Niroula, John and Sikder, Bejoy and Luo, Shisong and Fu, Kai and Rajput, Nitul S and Pranta, Ayan Biswas and Yadav, Pradyot and Zhao, Yuji and others},
  title={{Towards DTCO in high temperature GaN-on-Si technology: Arithmetic logic unit at 300° C and CAD framework up to 500° C}},
  booktitle={2023 IEEE Symposium on VLSI Technology and Circuits (VLSI Technology and Circuits)},
  pages={1--2},
  year={2023},
}

@article{2022gan,
  author={Yuan, Mengyang and Xie, Qingyun and Niroula, John and Chowdhury, Nadim and Palacios, Tomas},
  title={{GaN memory operational at 300° C}},
  journal={IEEE Electron Device Letters},
  volume={43},
  number={12},
  pages={2053--2056},
  year={2022},
}

@article{2016CMOS,
  author={Chu, Rongming and Cao, Yu and Chen, Mary and Li, Ray and Zehnder, Daniel},
  title={{An experimental demonstration of GaN CMOS technology}},
  journal={IEEE Electron Device Letters},
  volume={37},
  number={3},
  pages={269--271},
  year={2016},
}

@article{2019monolithic,
  author={Cui, Miao and Sun, Ruize and Bu, Qinglei and Liu, Wen and Wen, Huiqing and Li, Ang and Liang, Yung C and Zhao, Cezhou},
  title={{Monolithic GaN half-bridge stages with integrated gate drivers for high temperature DC-DC buck converters}},
  journal={IEEE Access},
  volume={7},
  pages={184375--184384},
  year={2019},
}

@article{2016highddd,
  author={Perrin, Remi and Quentin, Nicolas and Allard, Bruno and Martin, Christian and Ali, Marwan},
  title={{High-temperature GaN active-clamp flyback converter with resonant operation mode}},
  journal={IEEE Journal of emerging and selected topics in power electronics},
  volume={4},
  number={3},
  pages={1077--1085},
  year={2016},
}

@article{2022optimization,
  author={Shi, Ningping and Wang, Kejia and Zhou, Bing and Weng, Jiafu and Cheng, Zhiyuan},
  title={{Optimization AlGaN/GaN HEMT with field plate structures}},
  journal={Micromachines},
  volume={13},
  number={5},
  pages={702},
  year={2022},
}

@ARTICLE{2020A,
  author={Chowdhury, Nadim and Xie, Qingyun and Yuan, Mengyang and Cheng, Kai and Then, Han Wui and Palacios, Tomás},
  title={{Regrowth-Free GaN-Based Complementary Logic on a Si Substrate}}, 
  journal={IEEE Electron Device Letters}, 
  volume={41},
  number={6},
  pages={820-823},
  year={2020},
}

@article{2024ScAlN,
  author={Hasan, Md Tanvir and Liu, Jiangnan and Wang, Ding and Mondal, Shubham and Tanim, Md Mehedi Hasan and Yang, Samuel and Mi, Zetian},
  title={{Effect of temperature on the performance of ScAlN/GaN high-electron mobility transistor}},
  journal={Applied Physics Letters},
  volume={125},
  number={21},
  year={2024},
}

@article{2025investigation,
  author={Eisner, Savannah R and Liu, Yi-Chen and Naphy, Jared and Chen, Ruiqi and Rais-Zadeh, Mina and Senesky, Debbie G},
  title={{Investigation of InAlN/GaN Circular Transistors for Venus and Other High-Temperature Applications}},
  journal={IEEE Transactions on Electron Devices},
  year={2025},
  publisher={IEEE}
}

@article{2021epi,
  author={Sarkar, Ritam and Upadhyay, Bhanu B and Bhunia, Swagata and Pokharia, Ravindra S and Nag, Dhiman and Surapaneni, S and Lemettinen, Jori and Suihkonen, Sami and Gribisch, Philipp and Osten, Hans-J{\"o}rg and others},
  title={{Epi-Gd 2 O 3-MOSHEMT: A potential solution toward leveraging the application of AlGaN/GaN/Si HEMT with improved I ON/I OFF operating at 473 K}},
  journal={IEEE Transactions on Electron Devices},
  volume={68},
  number={6},
  pages={2653--2660},
  year={2021},
}

@article{2017inaln,
  author={Chapin, Caitlin A and Miller, Ruth A and Dowling, Karen M and Chen, Ruiqi and Senesky, Debbie G},
  title={{InAlN/GaN high electron mobility micro-pressure sensors for high-temperature environments}},
  journal={Sensors and Actuators A: Physical},
  volume={263},
  pages={216--223},
  year={2017},
}

@article{yuan2023enhancement,
  author={Yuan, Mengyang and Niroula, John and Xie, Qingyun and Rajput, Nitul S and Fu, Kai and Luo, Shisong and Das, Sagar Kumar and Iqbal, Abdullah Jubair Bin and Sikder, Bejoy and Isamotu, Mohamed Fadil and others},
  title={{Enhancement-mode GaN transistor technology for harsh environment operation}},
  journal={IEEE Electron Device Letters},
  volume={44},
  number={7},
  pages={1068--1071},
  year={2023},
}

@article{Applications2015,
  author={Watson, Jeff and Castro, Gustavo},
  title={{A review of high-temperature electronics technology and applications}},
  journal={Journal of Materials Science: Materials in Electronics},
  volume={26},
  pages={9226--9235},
  year={2015},
}

@article{Applications2024,
  author={Pradhan, Dhiren K and Moore, David C and Francis, A Matt and Kupernik, Jacob and Kennedy, W Joshua and Glavin, Nicholas R and Olsson III, Roy H and Jariwala, Deep},
  title={{Materials for high-temperature digital electronics}},
  journal={Nature Reviews Materials},
  pages={1--18},
  year={2024},
}

@book{Applications2023,
  author={Khanna, Vinod Kumar},
  title={{Extreme-temperature and harsh-environment electronics: physics, technology and applications}},
  year={2023},
  publisher={IOP Publishing}
}

@inproceedings{Hypersonic[14],
  author={Okojie, Robert S},
  title={{800 C Silicon Carbide (SiC) Pressure Sensors for Engine Ground Testing}},
  booktitle={{National Space and Missile Materials Symposium}},
  number={GRC-E-DAA-TN32392},
  year={2016}
}

@techreport{normann2006first,
  author={Normann, Randy Allen},
  title={{First high-temperature electronics products survey 2005.}},
  year={2006},
  institution={{Sandia National Laboratories (SNL), Albuquerque, NM, and Livermore, CA~…}}
}

@article{serp2014molten,
  author={Serp, J{\'e}r{\^o}me and Allibert, Michel and Bene{\v{s}}, Ond{\v{r}}ej and Delpech, Sylvie and Feynberg, Olga and Ghetta, V{\'e}ronique and Heuer, Daniel and Holcomb, David and Ignatiev, Victor and Kloosterman, Jan Leen and others},
  title={{The molten salt reactor (MSR) in generation IV: overview and perspectives}},
  journal={Progress in Nuclear Energy},
  volume={77},
  pages={308--319},
  year={2014},
}

@misc{guardian2024geothermal,
  author = {{The Guardian}},
  title = {{US aiming to ‘crack the code’ on deploying geothermal energy at scale}},
  howpublished = {\url{https://www.theguardian.com/environment/2024/apr/02/geothermal-energy-electricity}},
  year = {2024},
  note = {Accessed: 2024-05-02}
}

@article{valentian2022venus,
  author={Valentian, Dominique and Koppel, Christophe and Mairet, Philippe and Mairet, Loup},
  title={{Venus sample return mission revisited}},
  journal={Experimental Astronomy},
  volume={54},
  number={2},
  pages={597--616},
  year={2022},
}

@misc{bbc_haber,
  author = {{BBC Bitesize}},
  title = {{Haber process and Contact process - Higher - Making fertiliser - OCR Gateway - GCSE Chemistry (Single Science) Revision}},
  howpublished = {\url{https://www.bbc.co.uk/bitesize/guides/zxy9ng8/revision/5}},
  year = {2023},
}

@misc{jdpower_exhaust,
  author = {{J.D. Power}},
  title = {{How Hot Does A Car Exhaust Pipe Get?}},
  howpublished = {\url{https://www.jdpower.com/cars/shopping-guides/how-hot-does-a-car-exhaust-pipe-get}},
  year = {2024},
}

@article{Alpert2020SensitivityTemperatures,
    author = {Alpert, H S and Chapin, C A and Dowling, K M and Benbrook, S R and K{\"{o}}ck, H and Ausserlechner, U and Senesky, D G},
    title = {{Sensitivity of 2DEG-based Hall-effect sensors at high           temperatures}},
    journal = {Review of Scientific Instruments},
    number = {2},
    month = {2},
    pages = {25003},
    volume = {91},
    publisher = {American Institute of Physics},
    year = {2020},
}

@article{zhang2025wide,
  author={Zhang, Yuhao and Dong, Dong and Li, Qiang and Zhang, Richard and Udrea, Florin and Wang, Han},
  title={{Wide-bandgap semiconductors and power electronics as pathways to carbon neutrality}},
  journal={Nature Reviews Electrical Engineering},
  pages={1--18},
  publisher={Nature Publishing Group UK London},
  year={2025},
}

@article{mazumdar2019nanocrack,
  author={Mazumdar, Kaushik and Kala, Sanam and Ghosal, Aniruddha},
  title={{Nanocrack formation due to inverse piezoelectric effect in AlGaN/GaN HEMT}},
  journal={Superlattices and Microstructures},
  volume={125},
  pages={120--124},
  publisher={Elsevier},
  year={2019},
}

@article{AlN1,
  author={Zimmermann, Tom and Deen, David and Cao, Yu and Simon, John and Fay, Patrick and Jena, Debdeep and Xing, Huili Grace},
  title={{AlN/GaN insulated-gate HEMTs with 2.3 A/mm output current and 480 mS/mm transconductance}},
  journal={IEEE Electron Device Letters},
  volume={29},
  number={7},
  pages={661--664},
  publisher={IEEE},
  year={2008},
}

@article{AlN2,
  author={Harrouche, Kathia and Kabouche, Riad and Okada, Etienne and Medjdoub, Farid},
  title={{High performance and highly robust AlN/GaN HEMTs for millimeter-wave operation}},
  journal={IEEE Journal of the Electron Devices Society},
  volume={7},
  pages={1145--1150},
  publisher={IEEE},
  year={2019},
}

@article{ambacher2000two,
  author={Ambacher, Oliver and Foutz, B and Smart, Joseph and Shealy, JR and Weimann, NG and Chu, K and Murphy, M and Sierakowski, AJ and Schaff, WJ and Eastman, LF and others},
  title={{Two dimensional electron gases induced by spontaneous and piezoelectric polarization in undoped and doped AlGaN/GaN heterostructures}},
  journal={Journal of applied physics},
  volume={87},
  number={1},
  pages={334--344},
  publisher={American Institute of Physics},
  year={2000},
}

@article{hardy2017epitaxial,
  author={Hardy, Matthew T and Downey, Brian P and Nepal, Neeraj and Storm, David F and Katzer, D Scott and Meyer, David J},
  title={{Epitaxial ScAlN grown by molecular beam epitaxy on GaN and SiC substrates}},
  journal={Applied Physics Letters},
  volume={110},
  number={16},
  publisher={AIP Publishing},
  year={2017},
}

@article{2024pGaN,
  author={Ng, Yat H and Zheng, Zheyang and Zhang, Li and Liu, Ruizi and Chen, Tao and Feng, Sirui and Shao, Qiming and Chen, Kevin J},
  title={{p-GaN gate power HEMT heterostructure as a versatile platform for extremely wide-temperature-range (X-WTR) applications}},
  journal={Applied Physics Letters},
  volume={124},
  number={4},
  publisher={AIP Publishing},
  year={2024},
}

@article{lee2015temperature,
  author={Lee, In Hak and Kim, Yong Hyun and Chang, Young Jun and Shin, Jong Hoon and Jang, T and Jang, Seung Yup},
  title={{Temperature-dependent hall measurement of AlGaN/GaN heterostructures on Si substrates}},
  journal={Journal of the Korean Physical Society},
  volume={66},
  pages={61--64},
  publisher={Springer},
  year={2015}
}

@article{cuerdo2009high,
  author={Cuerdo, Roberto and Sillero, Eugenio and Romero, Mar{\'\i}a F{\'a}tima and Uren, Michael J and di Forte Poisson, Marie-Antoinette and Mu{\~n}oz, El{\'\i}as and Calle, Fernando},
  title={{High-temperature microwave performance of submicron AlGaN/GaN HEMTs on SiC}},
  journal={IEEE Electron Device Letters},
  volume={30},
  number={8},
  pages={808--810},
  publisher={IEEE},
  year={2009}
}

@article{ottaviani2020evaluation,
  author={Ottaviani, A and Palacios, P and Zweipfennig, T and Alomari, M and Beckmann, C and Bierb{\"u}sse, D and Wieben, J and Ehrler, J and Kalisch, H and Negra, R and others},
  title={{Evaluation of high-temperature high-frequency GaN-Based LC-oscillator components}},
  journal={IEEE Transactions on Electron Devices},
  volume={67},
  number={11},
  pages={4587--4591},
  publisher={IEEE},
  year={2020}
}

@article{akita2001high,
  author={Akita, Mitsutoshi and Kishimoto, Shigeru and Mizutani, Takashi},
  title={{High-frequency measurements of AlGaN/GaN HEMTs at high temperatures}},
  journal={IEEE Electron Device Letters},
  volume={22},
  number={8},
  pages={376--377},
  publisher={IEEE},
  year={2001}
}

@inproceedings{xue2024measurement,
  author={Xue, Hao and Storey, Craig and No{\"e}l, Jean-Paul and Griffin, Ryan},
  title={{Measurement and Modeling of GaN HEMTs Operating at 500° C}},
  booktitle={2024 IEEE Canadian Conference on Electrical and Computer Engineering (CCECE)},
  pages={532--536},
  organization={IEEE},
  year={2024}
}

@phdthesis{JohnThesis,
  author={Niroula, John},
  title={{Thermally Hardened RF GaN HEMTs in Extreme Environments}},
  school={Massachusetts Institute of Technology},
  year={2025}
}

@article{2025J,
  author={Niroula, John and Xie, Qingyun and Borujeny, Elham Rafie and Luo, Shisong and Oh, Minsik and Taylor, Matthew A and Zhao, Yuji and Palacios, Tom{\'a}s},
  title={{High Temperature AlGaN/GaN MISHEMT with W/AlON Gate Stack and I max> 1 A/mm at 500° C}},
  journal={IEEE Electron Device Letters},
  publisher={IEEE},
  year={2025}
}

@article{2025bootstrap,
  author={Xiong, Yixin and Visvkarma, Ajay K and Guan, Rian and Banner, Nathan S and Chiu, Chan-Wen and Zheng, Xiaojun and Mohney, Suzanne E and Jackson, Thomas N and Chu, Rongming},
  title={{GaN Bootstrapping Amplifier IC Operating at up to 800° C Temperature}},
  journal={IEEE Electron Device Letters},
  publisher={IEEE},
  year={2025}
}

@article{2025pGaN800,
  author={Visvkarma, Ajay Kumar and Gaona, Juan Nicolas Jimenez and Chiu, Chan-Wen and Xiong, Yixin and Huang, Yi-Shuo and Guan, Rian and Banner, Nathan S and Mohney, Suzanne E and Chu, Rongming},
  title={{p-GaN Gated HEMT with 770 I ON/I OFF Ratio Operating at 800° C}},
  journal={IEEE Electron Device Letters},
  publisher={IEEE},
  year={2025}
}

@article{Radiationcool,
  author={L{\'e}vesque, Luc},
  title={{Law of cooling, heat conduction and Stefan-Boltzmann radiation laws fitted to experimental data for bones irradiated by CO2 laser}},
  journal={Biomedical optics express},
  volume={5},
  number={3},
  pages={701--712},
  publisher={Optical Society of America},
  year={2014}
}

@ARTICLE{yuan2022,
  author={Yuan, Mengyang and Xie, Qingyun and Fu, Kai and Hossain, Toiyob and Niroula, John and Greer, James A. and Chowdhury, Nadim and Zhao, Yuji and Palacios, Tomás},
  title={{GaN Ring Oscillators Operational at 500 °C Based on a GaN-on-Si Platform}}, 
  journal={IEEE Electron Device Letters}, 
  volume={43},
  number={11},
  pages={1842-1845},
  year={2022},
}

@INPROCEEDINGS{Medjdoub2006,
  author={Medjdoub, F. and Carlin, J.-F. and Gonschorek, M. and Feltin, E. and Py, M.A. and Ducatteau, D. and Gaquiere, C. and Grandjean, N. and Kohn, E.},
  booktitle={2006 International Electron Devices Meeting}, 
  title={{Can InAlN/GaN be an alternative to high power/high temperature AlGaN/GaN devices?}}, 
  pages={1-4},
  doi={10.1109/IEDM.2006.346935},
  year={2006}}

@ARTICLE{Herfurth2013,
  author={Herfurth, Patrick and Maier, David and Lugani, Lorenzo and Carlin, Jean-Francois and Rosch, Rudolf and Men, Yakiv and Grandjean, Nicolas and Kohn, Erhard},
  journal={IEEE Electron Device Letters}, 
  title={{Ultrathin Body InAlN/GaN HEMTs for High-Temperature (600 $^{\circ} {\rm C}$) Electronics}}, 
  year={2013},
  volume={34},
  number={4},
  pages={496-498},
  doi={10.1109/LED.2013.2245625}}

@ARTICLE{Macron2011,
  author={Macron, D. and Kang, X. and Viaene, J. and Van Hove, M. and Srivastava, P. and Decoutere, S. and Mertens, R. and Borghs, G.},
  journal={Microelectronics Reliability}, 
  title={{GaN-based HEMTs tested under high temperature storage test}}, 
  year={2011},
  volume={51},
  number={9-11},
  pages={1717-1720}}

@ARTICLE{Liu2021environment,
  author={Liu, C. and Chen, Y. Q. and Liu, Y. and Lai, P. and He, Z. Y. and En, Y. F. and Wang, T. Y. and Huang, Y.},
  journal={IEEE Transactions on Electron Devices}, 
  title={{Effect of Atmosphere on Electrical Characteristics of AlGaN/GaN HEMTs Under Hot-Electron Stress}}, 
  year={2021},
  volume={68},
  number={3},
  pages={1000-1005},
  doi={10.1109/TED.2021.3049764}
}

@Article{Roccaforte2022sic,
  AUTHOR={Roccaforte, Fabrizio and Giannazzo, Filippo and Greco, Giuseppe},
  TITLE={{Ion Implantation Doping in Silicon Carbide and Gallium Nitride Electronic Devices}},
  JOURNAL={Micro},
  VOLUME={2},
  YEAR={2022},
  NUMBER={1},
  PAGES={23--53},
  URL={https://www.mdpi.com/2673-8023/2/1/2},
  ISSN={2673-8023},
  DOI={10.3390/micro2010002}
}

@article{Huang2022sic,
  author={Huang, Yuanchao and Wang, Rong and Zhang, Yiqiang and Yang, Deren and Pi, Xiaodong},
  title={{Compensation of p-type doping in Al-doped 4H-SiC}},
  journal={Journal of Applied Physics},
  volume={131},
  number={18},
  pages={185703},
  year={2022},
  month={05},
  issn={0021-8979},
  doi={10.1063/5.0085510},
  url={https://doi.org/10.1063/5.0085510},
  eprint={https://pubs.aip.org/aip/jap/article-pdf/doi/10.1063/5.0085510/16507991/185703\_1\_online.pdf}
}

@article{cui2022scaling,
  title={{Scaling behavior of InAlN/GaN HEMTs on silicon for RF applications}},
  author={Cui, Peng and Zeng, Yuping},
  journal={Scientific Reports},
  volume={12},
  number={1},
  pages={16683},
  year={2022},
  publisher={Nature Publishing Group UK London}
}

@article{sIc2024review,
  title={{Review and outlook on GaN and SiC power devices: Industrial state-of-the-art, applications, and perspectives}},
  author={Buffolo, M and Favero, D and Marcuzzi, A and De Santi, C and Meneghesso, G and Zanoni, E and Meneghini, M},
  journal={IEEE Transactions on Electron Devices},
  volume={71},
  number={3},
  pages={1344--1355},
  year={2024},
  publisher={IEEE}
}

@article{newaddedIrOx,
  title={{Degradation Analysis of InAlN/GaN Transistors
under Simulated Venus Surface Conditions}},
  author={Eisner, Savannah R. and Heussen, Zahra N. and Cordero, Sergio and Niroula, John and Naphy, Jared and Isamotu, Mohamed Fadil and Surdi, Harshad and Yuan, Mengyang and Xie, Qingyun and Nemanich, Robert J. and Palacios, Tomas and Rais-Zadeh, Mina and Hunter, Gary W. and Senesky, Debbie G.},
  journal={IEEE Transactions on Electron Devices},
  year={2025},
  publisher={IEEE}
}
